\RequirePackage{ifpdf}
\documentclass[12pt,a4paper,hyper]{article}
\pdfoutput=1

\usepackage{jheppub}
\usepackage{epsfig}
\usepackage{multirow}
\usepackage{url}
\usepackage{bbm}
\usepackage{graphicx}
\usepackage[utf8]{inputenc}
\usepackage[T1]{fontenc}
\usepackage{amssymb,amsfonts,amsmath,mathtools}

\usepackage{eurosym, xspace}
\usepackage[nottoc]{tocbibind}
\usepackage{subcaption, float}

\usepackage{fancyhdr}
\usepackage{makecell}
\usepackage{changepage}
\usepackage{multicol}
\usepackage{lastpage}
\usepackage{enumitem}
\usepackage{diagbox}
\usepackage{tabularx}
\usepackage{tabu}
\usepackage[usenames, dvipsnames, svgnames, table]{xcolor}

\font\cmss=cmss10
\font\cmsss=cmss10 at 7pt
\font\manual=manfnt

\newcommand{\bi}{\begin{itemize}}
\newcommand{\ei}{\end{itemize}}

\newcommand{\bea}{\begin{eqnarray}}
\newcommand{\eea}{\end{eqnarray}}
\newcommand{\be}{\begin{equation}}
\newcommand{\ee}{\end{equation}}
\newcommand{\ben}{\begin{eqnarray*}}
\newcommand{\een}{\end{eqnarray*}}
\newcommand{\bem}{\begin{pmatrix}}
\newcommand{\eem}{\end{pmatrix}}
\newcommand{\bl}{\begin{align}}
\newcommand{\el}{\end{align}}
\newcommand{\beg}{\begin{gather}}
\newcommand{\eeg}{\end{gather}}

\newcommand{\e}[0]{\mathrm{e}}
\newcommand{\I}[0]{\mathrm{i}}
\newcommand{\N}[0]{\ensuremath{\mathbb{N}}}
\newcommand{\Z}[0]{\ensuremath{\mathbb{Z}}}

\newcommand{\R}[0]{\ensuremath{\mathbb{R}}}
\newcommand{\C}[0]{\ensuremath{\mathbb{C}}}
\renewcommand{\Re}{\operatorname{Re}}

\newcommand{\dd}[0]{\mathrm{d}}
\newcommand{\pd}[0]{\partial}

\newcommand{\mc}[1]{{\mathcal{#1}}}

\DeclareMathOperator{\sign}{sign}
\newcommand{\ket}[1]{\mathinner{|#1 \rangle}}

\newcommand{\bracket}[2]{\mathinner{\langle #1 | #2 \rangle}}

\newcommand{\Mean}[1]{\left\langle #1 \right\rangle}
\newcommand{\abs}[1]{{|#1|}}

\newcommand{\com}[2]{[ #1, #2 ]}
\newcommand{\anticom}[2]{\{ #1, #2 \}}

\newcommand{\group}[1]{\mathrm{#1}}

\DeclareUnicodeCharacter{00A0}{~}
\DeclareUnicodeCharacter{202F}{~}

\newcommand{\what}[1]{\widehat{#1}}

\newenvironment{myenumerate}{
\begin{enumerate}
 \setlength{\itemsep}{1pt}
 \setlength{\parskip}{0pt}
 \setlength{\parsep}{0pt}}{\end{enumerate}}

\newcommand{\cH}{\mathcal{H}}

\newcommand{\cS}{\mathcal{S}}

\newcommand{\bC}{\ensuremath{\mathbb{C}}}

\newcommand{\bR}{\ensuremath{\mathbb{R}}}

\newcommand{\bZ}{\ensuremath{\mathbb{Z}}}

\newcommand{\IH}{\mathbb{H}}

\renewcommand{\a}{\alpha}
\renewcommand{\b}{\beta}
\renewcommand{\d}{\delta}

\newcommand{\m}{\mu}
\newcommand{\n}{\nu}

\renewcommand{\O}{\Omega}

\newcommand{\TrH[1]}{ {\raise -.5em
 \hbox{$\buildrel {\textstyle {\rm Tr } }\over
{\scriptscriptstyle \cH _ {#1}}$}~}}
\newcommand{\res[1]}{ {\raise -.5em
 \hbox{$\buildrel {\textstyle {\rm Res } }\over
{\scriptscriptstyle {#1}}$}}}

\renewcommand{\Re}{\mbox{Re}}

\definecolor{MyBlueDef}{rgb}{0,0,0.8}
\definecolor{MyPurpleDef}{rgb}{0.6,0.0,0.6}

\usepackage{marginnote}
\newcounter{draftcommentcnt}

\def\dbend{\lower3.5pt\hbox{\manual\char127}}

\def\IL{\relax{\rm I\kern-.18em L}}
\def\IH{\relax{\rm I\kern-.18em H}}

\def\rlx{\relax\leavevmode}

\def\ZZ{\rlx\leavevmode\ifmmode\mathchoice{\hbox{\cmss Z\kern-.4em Z}}
 {\hbox{\cmss Z\kern-.4em Z}}{\lower.9pt\hbox{\cmsss Z\kern-.36em Z}}
 {\lower1.2pt\hbox{\cmsss Z\kern-.36em Z}}\else{\cmss Z\kern-.4em
 Z}\fi}

\title{\center {Quantum Gravity from Timelike Liouville theory}}

\preprint{}

\author[1]{Teresa Bautista,}
\author[2,3,4]{Atish Dabholkar,}
\author[5]{Harold Erbin}

\affiliation[1]{Max Planck Institute for Gravitational Physics (Albert Einstein Institute) \\
	Mühlenberg 1, D-14476 Potsdam, Germany}
\affiliation[2]{International Centre for Theoretical Physics\\
	Strada Costiera 11, Trieste 34151 Italy}
\affiliation[3]{Sorbonne Université, UPMC Univ Paris 06\\
	\it UMR 7589, LPTHE, F-75005, Paris, France}
\affiliation[4]{CNRS, UMR 7589, LPTHE, F-75005, Paris, France}
\affiliation[5]{Ludwig--Maximilians--Universität\\ 
	München Theresienstraße 37, 80333 München, Germany}

\emailAdd{teresa.bautista@aei.mpg.de}
\emailAdd{atish@ictp.it}
\emailAdd{harold.erbin@physik.lmu.de}

\abstract{
\vspace{3mm}

A proper definition of the path integral of quantum gravity has been a long-standing puzzle because the Weyl factor of the Euclidean metric has a wrong-sign kinetic term. We propose a definition of two-dimensional Liouville quantum gravity with cosmological constant using conformal bootstrap for the timelike Liouville theory coupled to supercritical matter. We prove a no-ghost theorem for the states in the BRST cohomology. We show that the four-point function constructed by gluing the timelike Liouville three-point functions is well defined and crossing symmetric (numerically) for external Liouville energies corresponding to \textit{all} physical states in the BRST cohomology with the choice of the Ribault--Santachiara contour for the internal energy. 
\vspace{5mm}
}


\begin{document}
\maketitle

\newpage

\section{Introduction}
\label{sec:intro}

The path integral of Euclidean quantum gravity is ill-defined because the Weyl factor of the Euclidean metric has a wrong-sign kinetic term~\cite{Gibbons:1978ac}. It has been a long-standing puzzle how to make sense of such a theory. 
In this work, we address this problem for two-dimensional gravity with a cosmological constant, where the dynamics of the Weyl factor is described by the exactly-solvable Liouville conformal field theory~\cite{Polyakov:1981:QuantumGeometryBosonic}. This makes it possible to use the non-Lagrangian conformal bootstrap rather than the path integral to define quantum gravity coupled to conformal matter.

A simpler version of the problem is encountered in defining the Euclidean worldsheet path integral of critical bosonic string theory in $26$-dimensional Minkowski spacetime. The path integral for the timelike boson has a wrong-sign kinetic term and is ill-defined. It is nevertheless Gaussian. One can define it first for the spacelike boson with the right-sign kinetic term to compute all correlations and then analytically continue the result to the physical Minkowski energy. The analytic continuation is greatly simplified by the fact that (spacetime) energy is conserved as a consequence of the shift symmetry of the free timelike boson. However, even in this simple case, the analytic continuation of the \textit{integrated} correlations is subtle and requires a proper  `$\I\epsilon $ prescription' to make sense of the resulting worldsheet integrals~\cite{Witten:2013:FeynmaniEpsilon, Sen:2017:EquivalenceTwoContour}. 

The timelike boson of the critical string could be thought of as the Liouville mode of worldsheet gravity~\cite{Das:1989:QuantizationLiouvilleMode} coupled to non-critical matter of $25$ spacelike bosons with the cosmological constant set to zero. 
With the addition of the cosmological constant, one obtains a non-trivial interacting quantum field theory with a non-polynomial, exponential interaction. The action is no longer shift-symmetric and energy is no longer conserved. This makes the problem of analytic continuation much more severe already at the level of the \textit{unintegrated} correlation functions. One can hope that this solvable model is nevertheless tractable and holds some general lessons that can be extended to higher-dimensional gravity. 

With this perspective, we consider timelike Liouville theory coupled to supercritical matter. Timelike Liouville theory as a conformal field theory is by now well understood~\cite{Zamolodchikov:2005:ThreepointFunctionMinimal, Kostov:2006:BulkCorrelationFunctions, Kostov:2007:NonRational2DQuantum-1, Kostov:2007:NonRational2DQuantum-2, Ribault:2015:LiouvilleTheoryCentral, Ikhlef:2016:ThreepointFunctionsc}. The spectrum of operators and structure constants that satisfy the bootstrap equations are known. The path integral corresponding to the three-point function computed from the bootstrap has been studied in~\cite{Harlow:2011:AnalyticContinuationLiouville, Giribet:2012:TimelikeLiouvilleThreepoint}. To construct the corresponding quantum theory of gravity we need to impose diffeomorphism invariance after coupling the Liouville theory to matter. Our guiding principles are BRST invariance to implement the gauge symmetry, a no-ghost theorem to ensure positive norm for physical states, and a proper analytic continuation of correlations that ensures crossing symmetry.
Consistent with these physical requirements, 
we propose a definition of Euclidean quantum gravity on a 2-sphere, or equivalently on a plane, with a non-zero positive cosmological constant. 

We do not consider higher genus Riemann surfaces and the issue of possible tachyonic divergences at higher loops. Nor do we analyze the question of the $i\epsilon$ prescription for integrated correlations. These problems are left for the future. 
We summarize below our two main results.
 \vfill\eject
\begin{enumerate}
\item
\textit{A no-ghost theorem for the states in the BRST cohomology} 

We compute the BRST cohomology by generalizing the derivation 
in~\cite{Bouwknegt:1992:BRSTAnalysisPhysical} to include a timelike Coulomb gas boson. Hermiticity of the matter sector and the presence of the cosmological term truncate the cohomology further to a subset. We prove the {no-ghost theorem} by showing that all states in this physical Hilbert space have positive norm. 
\item
\textit{Crossing-symmetric correlations for all physical energies}

The physical spectrum of ghost-free BRST-invariant states allows for both real and imaginary Liouville energy. This is larger than the spectrum used in the conformal bootstrap~\cite{Ribault:2015:LiouvilleTheoryCentral} which includes only imaginary energies. It is necessary to find a proper analytic continuation so that four-point functions can be defined for all physical energies. 
A key new ingredient in our analysis is the observation that for a physically sensible theory it is not necessary to identify the external and internal spectra.
We show that the four-point function constructed by gluing the timelike three-point functions is well defined and crossing symmetric with the choice of the Ribault--Santachiara contour for the internal energy for \textit{all} physical states in the BRST cohomology. Crossing symmetry follows from the numerical analysis performed by Ribault and Santachiara.
\end{enumerate}

Our unusual prescription for the analytic continuation is inspired by recent work in string field theory~\cite{Pius:2016:CutkoskyRulesSuperstring, Sen:2017:EquivalenceTwoContour, deLacroix:2017:ClosedSuperstringField}. One of the features of string field theory scattering amplitudes is that the vertices in momentum space are exponentially damped for Euclidean energies, reflecting the soft ultraviolet behavior of strings. This implies though that vertices make an exponentially growing contribution to the contour at infinity for real energy. As a result, the integration contour along the imaginary energy axis cannot be analytically Wick-rotated to a contour along the real energy axis. One can nevertheless perform an analytic continuation by analytically-\textit{deforming} the contour appropriately to avoid poles but holding the ends at imaginary infinity fixed without rotating the whole contour. This prescription for the analytic continuation has been shown to yield physically sensible results in string field theory and can also be used consistently for standard QFTs~\cite{Pius:2016:CutkoskyRulesSuperstring, Sen:2016:UnitaritySuperstringField, Sen:2016:OneLoopMass, deLacroix:2017:ClosedSuperstringField, Pius:2018:UnitarityBoxDiagram, deLacroix:2018:AnalyticityCrossingSymmetry}.

In Liouville theory one encounters a similar problem already at \textit{tree level} for the unintegrated correlation functions because the timelike three-point function~\cite{Zamolodchikov:2005:ThreepointFunctionMinimal, Kostov:2006:BulkCorrelationFunctions, Kostov:2007:NonRational2DQuantum-1, Kostov:2007:NonRational2DQuantum-2} and conformal blocks exhibit exponential growth in regions of the complex energy plane corresponding to large negative conformal dimensions.
Examination of the poles in the integrand for the four-point correlation function reveals a  simplification compared to string field theory in that  the poles in fact  do not depend on the external states and do not move as the external energies are analytically continued. Consequently, it is not even necessary to deform the contour of integration to analytically continue the answer. Using the numerical code released with~\cite{Ribault:2015:LiouvilleTheoryCentral} we have explicitly checked that the resulting four-point functions are indeed well defined and crossing symmetric for all physical energies. This is one of our main conclusions.

Our results obtained for a Euclidean worldsheet with a spherical topology may have implications for the de Sitter spacetime if one can analytically continue the worldsheet time  to obtain Lorentzian signature on base space. Two-dimensional de Sitter-like solutions can be used as models for expanding cosmologies~\cite{Polchinski:1989:TwodimensionalModelQuantum,Dabholkar:2016:QuantumWeylInvariance, Bautista:2016:QuantumCosmologyTwo}. 
It was shown in~\cite{Dabholkar:2016:QuantumWeylInvariance, Bautista:2016:QuantumCosmologyTwo} that inclusion of nonlocal quantum corrections modifies the barotropic index of vacuum energy. This leads to the dilution of vacuum energy that is consistent with momentum conservation and slows down the de Sitter expansion. This interesting mechanism depends on a choice of the initial vacuum and a particular choice of the `physical' metric. In general, there is no single choice of the metric for renormalizing all operators in the effective action that is mutually local, because of the gravitational dressings. It is a rather subtle question to determine which metric and which vacuum should be regarded as physical. 
The semiclassical limit of Liouville gravity that we construct in this paper can perhaps shed some light on these issues.

Earlier work on two-dimensional gravity using matrix models concerns spacelike Liouville theory coupled to subcritical matter. Since spacelike Liouville theory has the right-sign kinetic term, it is not quite like the Weyl factor of the metric in higher-dimensional gravity. Another advantage of timelike Liouville theory is that it admits a semiclassical limit with unitary matter, which is amenable to a path integral treatment~\cite{Harlow:2011:AnalyticContinuationLiouville, Giribet:2012:TimelikeLiouvilleThreepoint}. In these respects timelike Liouville theory coupled to unitary matter is a better model for higher-dimensional gravity.

Another motivation for the present work is the $AdS/CFT$ correspondence. If one wishes to extend holography beyond the large $N$ classical gravity limit, it is necessary to be able to define the path integral of supergravity in the bulk-$AdS$ spacetime. It has been possible to apply localization techniques to the formal supergravity path integral in bulk $AdS$ to compute finite $N$ or finite Planck length effects corresponding to genuine quantum gravity corrections. Localization often reduces the path integral to a finite integral. To obtain sensible answers in agreement 
with the boundary $CFT$, it is necessary to rotate the contour of integration for the zero mode of the Weyl factor of the metric~\cite{Dabholkar:2010uh,Dabholkar:2011ec, Dabholkar:2014wpa} in the final stages of the computation.\footnotemark{}
\footnotetext{%
	For example, in the context of $AdS_{2}/CFT_{1}$ holography, the Hardy--Ramanujan--Rademacher expansion of the quantum degeneracies of the black hole from the boundary involves generalized Bessel functions. Localization in the bulk reproduces the correct Bessel integral but the correct Bessel contour corresponds to rotating the integration over the zero mode of the Weyl factor from the real axis to the imaginary axis~\cite{Dabholkar:2011ec}.
}%
It would be more satisfactory to have an \textit{a priori} definition of the supergravity path integral which can justify such an analytic continuation from first principles. An advantage of the holographic setup is that comparison with the boundary CFT can be used as an independent check for the answer. To discuss an asymptotically $AdS$ worldsheet, it would be necessary to extend our results to worldsheets with a boundary. This is an interesting problem for the future. 

One can analyze the timelike Liouville theory using minisuperspace analysis~\cite{Strominger:2002:OpenStringCreation, Gutperle:2003:TimelikeBoundaryLiouville, Strominger:2003:CorrelatorsTimelikeBulk, Schomerus:2003:RollingTachyonsLiouville, Fredenhagen:2003:MinisuperspaceModelsSbranes, McElgin:2008:NotesLiouvilleTheory}.  In particular, one may hope to compute the spectrum and correlation functions of light states from the minisuperspace wave functions. 
However, this analysis is plagued with difficulties  and we were not able to draw any clear conclusions. We review some of the issues in  \S\ref{sec:history} and describe our analysis briefly in \S\ref{sec:minisuperspace} and at more length in~\cite{Bautista:BRSTCohomologyTimelike}. 
Given these difficulties, we  renounce  the minisuperspace altogether and use BRST invariance and crossing symmetry as our guiding principles.

The outline of the paper is as follows. We 
review generalities about Liouville field theory in~\S\ref{sec:qg} and the conformal field theory of timelike Liouville in~\S\ref{sec:TLT}. 
In \S\ref{sec:brst} we present a derivation of the BRST cohomology for timelike Liouville theory coupled to $D\geq 25$ free bosons and prove the no-ghost theorem. 
In \S\ref{sec:ana} we present the prescription for defining the correlation functions explaining the analogy with string field theory (which is reviewed briefly in \S\ref{sec:sft}). 
We conclude in \S\ref{sec:discussion} with a discussion of possible extensions.
Finally, few comments on the minisuperspace analysis are given in \S\ref{sec:minisuperspace} and a brief history of timelike Liouville is provided in \S\ref{sec:history}.

\section{Quantum gravity in two dimensions}
\label{sec:qg}

Our starting point is the path integral for the two-dimensional Euclidean metric $g_{\m\n}$ coupled to $D$ free scalar fields $\{ Y^I \}$ with central charge $c_m=D$:
\begin{equation}
	Z
		= \int
		\frac{1}{V_{\text{diff}}} \, \mathcal{D} g_{\mu\nu}
 \, \mathcal{D} Y^I \, \e^{ - S_m[g, Y^I] - S_g[g] } \, ,
\end{equation}
where $V_{\text{diff}}$ is the volume of the diffeomorphism group. The matter action is given by
\begin{equation}
	S_{m} = \frac{1}{4\pi}\sum_{I=1}^{D}\int d^2 x \sqrt{g} \, g^{\m\n} \nabla_{\m} Y^{I}\, \nabla_{\n} Y^{I} \, , 
\end{equation}
and the classical gravitational action is given by
\begin{equation}
S_{g} = \frac{1}{4\pi}\int d^2 x \sqrt{g} \, R[g] \, + \, 
 \mu_0 \int d^2 x \sqrt{g}
\end{equation}
where $\mu_0$ is the bare cosmological constant. The Einstein--Hilbert term is topological and will play no role since we will restrict ourselves to the spherical topology. Our goal is to make sense of this path integral.

\subsection{Generalities about Liouville gravity}

It is convenient to introduce a Weyl compensator $\Omega(x)$ and a fiducial metric $\bar g_{\mu\nu}(x)$ to write \begin{equation}
	g_{\mu\nu}(x) = \e^{2 \Omega(x)} \bar g_{\mu\nu}(x) \, .
\end{equation}
This split is arbitrary and the metric is clearly invariant under a `\textit{fiducial} Weyl transformation'
\begin{equation}
	\label{emergent-Weyl-transf}
	\bar g_{\mu\nu} \rightarrow \bar g_{\mu\nu} \, \e^{2\sigma(x)}
	\qquad \text{and} \qquad
	\Omega(x) \rightarrow \Omega(x) - \sigma(x) \, .
\end{equation}
We have effectively introduced a new scalar degree of freedom but at the same time we have enlarged the gauge symmetry so that the classical dynamics does not change. 
Fadeev--Popov gauge fixing of the gravitational measure yields
\begin{equation}
	\label{eq:gauge-fixing}
	\frac{1}{V_{\text{diff}}} \, \mathcal{D} g_{\mu\nu}
		= \mathcal{D} \Omega \, \mathcal{D} (b,c) \, \e^{- S_{\text{gh}}[g,b,c]} \, 
\end{equation} 
where the ghost action reads
\begin{eqnarray}
	\label{eq:action-ghosts}
	S_{\text{gh}}[g, b, c]
		= \frac{1}{4\pi} \int d^2 x \sqrt{g} \, b_{\mu\nu} \left[ 2\, \nabla^{(\mu} c^{\nu)} - g^{\mu\nu} \nabla_\rho c^\rho \right] \, .
\end{eqnarray}

Note that the cosmological constant is dimensionful and the original theory is \textit{not} invariant under the Weyl transformations of the physical metric. 
The action for matter and ghosts on the other hand is Weyl invariant, $S[g,Y^I,b,c] = S[\bar g, Y^I,b,c]$, but their measure is not. The anomalous Weyl variations of the measures can be deduced from the Weyl anomaly~\cite{Polyakov:1981:QuantumGeometryBosonic}: 
\begin{eqnarray}
	\label{Jacobian-matter-measure}
	\mathcal{D} Y^I &=& \e^{- \frac{c_m}{6} 
	\, S_{WZ}[\bar g, \Omega]} \, \overline{\mathcal{D}} Y^I \, , \\
	\mathcal{D} (b,c) &=& \e^{- \frac{c_{\text{gh}}}{6} \, S_{WZ} [\bar g, \Omega]} \, \overline{\mathcal{D}} (b,c) \, ,
	\qquad
	c_{\text{gh}} = - 26,
\end{eqnarray}
where
\begin{equation}
	S_{WZ}[\bar g, \Omega] = \frac{1}{4\pi} \int \dd^2 x \sqrt{\bar g} \left[ -(\overline \nabla \Omega)^2 - \bar R \, \Omega \right]
\end{equation} 
is the Wess--Zumino action corresponding to the Weyl anomaly.
The bar over the measure indicates that it is defined with respect to the fiducial metric, for example, using the norm
\begin{equation}
(\d Y, \d Y) = \int d^2 x \sqrt{\bar g}\, \d Y(x) \, \d Y(x) \, .
\end{equation}
The path integral then factorizes as $Z = Z_{m} \, Z_{\text{gh}} \, Z_\Omega$ with
\begin{equation}
	Z_{m}
		= \int \overline{\mathcal{D}} Y^I\, \e^{- S_m[\bar g, Y^I]} \, ,
	\quad
	Z_{\text{gh}}
		= \int \overline{\mathcal{D}} (b,c) \, \e^{- S_{\text{gh}}[\bar g, b, c]} \, ,
	\quad
	\label{Omega-partition-function-Weyl-measure}
	Z_{\Omega}
		= \int \mathcal{D} \Omega \, \e^{\frac{c_L}{6} S_{L}[\bar g, \Omega]} \, ,
\end{equation}
where 
\begin{equation}
	S_{L}[\bar g, \O] = \frac{1}{4\pi} \int \dd^2 x \sqrt{\bar g} \, \left[ - (\overline \nabla \O)^2 - \bar R \O + 4\pi \mu \, \e^{2 \O} \right] \, 
\end{equation} 
is the classical Liouville action with the renormalized cosmological constant $\mu$, and the
classical central charge $c_{L}$ given by
\begin{equation}\label{classicalc}
	c_{L} := - 6q^{2} := 26 - c_m \, .
\end{equation} 

The measure $\mathcal{D} \Omega$ for the Weyl factor is still defined with respect to the physical metric and is invariant under fiducial Weyl transformations. It is well known that one can convert it to the canonical measure $\overline{\mathcal{D}} \Omega$ that is invariant under field translations. The only effect of the change of measure is to renormalize the Liouville action in a manner that is invariant under the fiducial Weyl symmetry~\cite{David:1988:ConformalFieldTheories, Distler:1989:ConformalFieldTheory, Mavromatos:1989:RegularizingFunctionalIntegral, DHoker:1990:2DQuantumGravity, DHoker:1991:EquivalenceLiouvilleTheory}. 
By further rescaling $\chi = q\,\Omega $
we obtain the canonically-normalized action
\begin{equation}
	\label{timelike-Liouville-action}
	S_{tL}[\bar g, \chi] = \frac{1}{4\pi} \int \dd^2 x \sqrt{\bar g} \, \left[ - (\overline \nabla \chi)^2 - q \bar R \chi + 4\pi \mu \, \e^{2 \beta \chi} \right] \, .
\end{equation} 
Because the kinetic 
term has a wrong-sign, we refer to this action as \textit{timelike} Liouville. We have assumed $q$ is real, so the rescaling does not change the sign of the kinetic term. The unknown parameter $\b$ will be determined shortly.

A well-known and remarkable fact is that the Liouville field behaves like a free field for the purposes of renormalization~\cite{DHoker:1990:2DQuantumGravity, DHoker:1991:EquivalenceLiouvilleTheory}. Thus, the classical central charge \eqref{classicalc} receives an additional quantum contribution from the Liouville field as if for a free scalar field:
\begin{equation}\label{cq}
	c_L := 1 - 6 q^2 \, .
\end{equation} 
For $q$ real we have $c_{L}\leq 1$. 
Moreover, free-field normal ordering removes all ultraviolet divergences. In particular, the dimension of the composite exponential operator is, as for a free field with background charge:
\begin{equation}
	\Delta\big(\e^{2 \beta \chi}\big)
		= \beta (q + \beta) \, .
\end{equation} 
The requirement of 
invariance under the fiducial Weyl invariance then implies
\begin{equation}
	\label{eq:q-beta}
	q = \frac{1}{\beta} - \beta \, , 
\end{equation} 
where the second piece can be interpreted as a quantum correction to the classical value $q = \b^{-1}$. 

\subsection{BRST quantization}

The net result of the introduction of the Weyl compensator and the gauge fixing is that both the matter and the gravity sectors are now described by a (fiducial)-Weyl-invariant action with \textit{canonical} measure. Ghosts and matter are free theories and they can be dealt with as in critical string theory. 
For timelike Liouville, we need to make sense of the quantum theory that corresponds to the Liouville path integral above with canonical measure and $q \in \R$, that is, when $c_L \le 1$ or $c_m \ge 25$ taking into account the quantum correction in \eqref{cq} and \eqref{csum}. 

Diffeomorphism invariance of the theory in the original field variables $\{g_{\mu\nu} \}$ is now equivalent to diffeomorphism plus fiducial Weyl invariance in the new variables $\{\bar g_{\mu\nu} \, , \Omega \}$. For a diffeomorphism invariant theory, Weyl invariance is equivalent to conformal invariance. Our construction using the bootstrap will be manifestly conformally invariant. 
To implement diffeomorphism invariance, we use BRST quantization. The BRST quantization of this theory is similar to the BRST quantization of the critical string or of two-dimensional gravity corresponding to spacelike Liouville coupled to subcritical matter. There is extensive literature on the topic which can be easily adapted to the case at hand of two-dimensional gravity corresponding to timelike Liouville coupled to supercritical matter. We review the relevant background in \S\ref{sec:brst}. The BRST operator $Q_B$ is nilpotent iff the total central charge vanishes as in the case of critical string theory. This implies
\begin{equation}
\label{csum}
	c_{m} - 26 + 1 - 6q^{2} = 0 \, , \qquad \text{or} \qquad q^2 := \frac{c_m - 25}{6}.
\end{equation}

Given a nilpotent BRST operator, there are two steps to construct a gauge-invariant two-dimensional gravity. We first need to determine the gauge-invariant spectrum of physical operators and prove a no-ghost theorem for the operators in the BRST cohomology. This is described in \S\ref{sec:brst}. We then need to give a prescription to compute correlation functions for all these operators. 
In flat fiducial metric, the integrated BRST-invariant operators are of the form
\begin{equation}
	\mc V(\alpha, k) = \int \dd^2 z \, V_\alpha(z, \bar z) V_m(z, \bar z; k),
\end{equation} 
where $V_m$ and $V_\alpha$ are respectively operators in the matter and Liouville sectors with quantum numbers $k$ and $\alpha$. Correlations in the matter sector are straightforward for free scalars. Computation of the correlation functions in the Liouville sector can be achieved using conformal bootstrap and after an appropriate analytic continuation as discussed in \S\ref{sec:TLT:corr} and \S\ref{sec:ana}. 

We comment briefly on the role of the $b, c$ ghosts since the vertex operators above are defined without any $c$ ghosts. 
The states in the BRST cohomology are naturally expressed as unintegrated operators on the $\ket{\downarrow}$ ghost vacuum, which is equivalent to the traditional $\bar c(\bar z) c(z)$ insertion.
If all vertex operators are given in this representation, the amplitude is obtained by integrating over the marked moduli space of genus-$g$ Riemann surfaces with $n$ punctures~\cite{Polchinski:2005:StringTheory-1}.
The definition of the measure includes some $b$-ghost insertions required to soak up the zero-modes.
In the case of the sphere, combining them with all but three $c$-ghost insertions leads to the standard formulation of amplitudes in critical string theory.
These remaining ghosts arise from the gauge fixing of $\group{SL}(2, \C)$.
Hence, with the no-ghost theorem proven in \S\ref{sec:brst:noghost}, all ghosts which appear are related to the measure of integration over the moduli space and not to negative-norm states.

\subsection{Timelike versus spacelike Liouville}

One can allow the parameter $q$ to be complex to explore the parameter space. 
From the perspective of the path integral for two-dimensional gravity it is natural to require that in appropriate real field variables the action be real. 
This singles out two regimes which we refer to as \textit{timelike} and \textit{spacelike} Liouville\footnotemark{}
\footnotetext{%
	From the CFT perspective, one may relax the requirement of reality of the action and regard the spacelike kinetic term with $c_L \le 1$ and the timelike kinetic with $c_L \ge 25$ as two additional independent possibilities~\cite{Ribault:2015:LiouvilleTheoryCentral}.
}%
using the target space terminology\footnotemark{}
\footnotetext{%
	Even though our main motivation is  $2d$-gravity in its own right, the target space interpretation may have interesting applications, for example, in the context of supercritical string theory or S-branes~\cite{Silverstein:2001xn, Strominger:2002:OpenStringCreation, Strominger:2003:CorrelatorsTimelikeBulk}.
}%
of critical strings:
\begin{myenumerate}
\item
Timelike : $c_L \le 1$ and $q, \b \in \R$.
\item
Spacelike : $c_L \ge 25$ and $q, \b \in \I \R$.
\end{myenumerate}
We refer to the corresponding regimes of gravity as `timelike gravity' and `spacelike gravity'.
For $q$ pure imaginary, the action \eqref{timelike-Liouville-action} can be made pure real by an `analytic continuation':
\begin{equation}
	\label{eq:analytic-cont-chi-q}
	\phi = \I \chi,
	\qquad
	Q = \I q,
	\qquad
	b = - \I \beta \, ,
\end{equation} 
to obtain the canonical spacelike Liouville action
\begin{equation}
	\label{spacelike-Liouville-action}
	S_{sL}[\bar g, \phi] = \frac{1}{4\pi} \int \dd^2 x \sqrt{\bar g} \, \left[ (\bar \nabla \phi)^2 + Q \bar R \phi + 4\pi \mu \, \e^{2 b \phi} \right] \, .
\end{equation} 
This action describes a Coulomb gas with background charge $Q$ deformed by the exponential interaction corresponding to the cosmological constant, with
\begin{equation}
Q = \frac{1}{b} + b \, 
\end{equation}
to ensure that the cosmological constant operator has conformal dimension $(1, 1)$.
The central charge is 
\begin{equation}
c_{L} = 1 + 6 Q^{2} \geq 25 \qquad \text{or} \qquad Q \geq 2 \, .
\end{equation}
The regime $c_L \in (1, 25)$ and $Q \in (0, 2)$ is also possible from the point of the CFT but requires complex $b$ and we do not consider it here.

One can rewrite the timelike action \eqref{timelike-Liouville-action} by rescaling $\chi \rightarrow q \chi$ so that there is an overall factor of $q^{2}$ which can then be interpreted as $1/\hbar$. Similar rescaling for the spacelike theory tells us that the semiclassical limit in the two regimes corresponds to 
\begin{equation}
q \gg 1 \, \qquad \text{and} \qquad Q \gg 1 \, .
\end{equation}
The parameters $\b$ and $b$ are sometimes referred to as the Liouville coupling constants. We equivalently define the semiclassical limit as
\begin{equation}
\b \ll 1 \, \qquad \text{and} \qquad b \ll 1 \, . 
\end{equation}

While the na\"ive analytic continuation \eqref{eq:analytic-cont-chi-q} at the level of the classical action is straightforward, it is far from clear how to analytically continue the integration cycle in the field space for the corresponding path integral. Generically, one would expect that the two semiclassical regimes could be separated by Stokes lines~\cite{Harlow:2011:AnalyticContinuationLiouville}. Indeed, this is the crux of the matter and has been the source of much of the difficulties in defining the timelike Liouville theory.
One of our main results is to obtain a definition of the timelike quantum gravity using conformal bootstrap and BRST invariance which would agree with the path integral defined using the action \eqref{timelike-Liouville-action} in the semiclassical regime. We obtain a physically-sensible definition for  timelike gravity; however it \textit{cannot} be obtained by a na\"ive analytic continuation from spacelike gravity.

\section{Liouville conformal field theory}
\label{sec:TLT}

In the conformally-flat gauge, the line element is
\begin{equation} \label{cgauge}
\bar g_{\mu\nu} \dd x^\mu \dd x^\nu = \abs{\dd z}^2 \, .
\end{equation} 
The timelike Liouville action \eqref{timelike-Liouville-action} takes the form\footnotemark{}
\footnotetext{%
	In this gauge, one must regulate the theory by restricting the plane to a finite disk with the radius as the IR regulator~\cite{Harlow:2011:AnalyticContinuationLiouville}. This introduces boundary terms in the action which we omit for simplicity.
}%
\begin{equation}
	\label{flat-timelike-Liouville-action}
	S_{tL}[\chi] = \frac{1}{2\pi} \int \dd^2 z \,\left[- \partial \chi \bar\partial \chi + \pi\mu \, \e^{2 \beta \chi} \right] \, .
\end{equation}
The equation of motion is 
\begin{equation}
	\partial \bar\partial \chi = - \pi \mu \,\beta \, \e^{2\beta \chi} \, . 
\end{equation}
The holomorphic momentum tensor is
\begin{equation}
	T(z) = (\partial \chi)^2 - q \, \partial^2 \chi \, ,
\end{equation} 
which can be derived from the covariant form of the action \eqref{timelike-Liouville-action}.
This corresponds to the momentum tensor of a Coulomb gas and shows that the Liouville theory is indeed a CFT (up to redundant operators which vanish upon using equations of motion).
The Liouville action is invariant under diffeomorphisms but shifts under the fiducial Weyl transformations by a field independent term. The equations of motion are therefore invariant under conformal transformations
\begin{equation}
	z' = w(z) \, 
	\qquad \text{and} \qquad
	\chi'(z', \bar{z}') = \chi(z, \bar z) - \frac{q}{2}\,\log \left(\frac{\partial w}{\partial z} \frac{\partial \bar w}{\partial \bar z} \right) \, ,
\end{equation}
which can be thought of as a diffeomorphism of the  flat space combined with a fiducial Weyl transformation. The latter brings the fiducial metric back to the flat metric as in the conformal gauge \eqref{cgauge}. 

In what follows, we treat Liouville conformal field theory using the non-Lagrangian conformal bootstrap approach to construct a theory whose semiclassical limit corresponds to the path integral defined using the action above.

\subsection{Conformal bootstrap for Liouville \label{Conformal}}

Within the framework of the BPZ conformal bootstrap~\cite{Belavin:1984:InfiniteConformalSymmetry} a conformal field theory is completely specified by the spectrum of primary operators and their three-point structure constants satisfying the crossing symmetry constraints. Within this framework, the quantum Liouville theory can be defined \textit{axiomatically} by the following two requirements:\footnotemark{}
\footnotetext{%
	See~\cite{Teschner:2001:LiouvilleTheoryRevisited, Nakayama:2004:LiouvilleFieldTheory, Pakman:2006:LiouvilleTheoryAction, Zamolodchikov:2007:LecturesLiouvilleTheory, Ribault:2014:ConformalFieldTheory} for reviews.
	The  $c_L \ge 25$ theory has  been constructed also using probabilistic methods~\cite{David:2016:LiouvilleQuantumGravitySphere, Kupiainen:2016:ConstructiveLiouvilleConformal, Rhodes:2016:LectureNotesGaussian}.
}%
\begin{myenumerate} 
\item The spectrum of dimensions $\Delta \in \bR$ is continuous. For each dimension, the multiplicity of the Virasoro representation is either one or zero.
That is, for a given dimension, either there is exactly one primary field or none. 
It is convenient to label the primary fields by the `\textit{Liouville charge}' which we denote by $\a$ for timelike Liouville and $a$ for spacelike Liouville ($\a, a \in \bC$) with $\alpha=\I a$ under the na\"ive analytic continuation \eqref{eq:analytic-cont-chi-q}. The spectrum of primary operators is correspondingly denoted by $\{V_{\a} \}$ and $\{V_{a} \}$ as we describe in more detail in the next subsection. 
\item The correlation functions are meromorphic functions of the external Liouville charges and the Liouville coupling constant. We denote the three-point structure constants by $\widehat C(\a_{1} , \a_{2}, \a_{3})$ and $ C(a_{1} , a_{2}, a_{3})$ respectively for timelike and spacelike Liouville with an implicit dependence on the couplings $\b$ and $b$. 
\end{myenumerate} 

The operator product expansion (OPE) for timelike Liouville takes the form
\begin{equation}
	\label{eq:generic-ope}
	V_{\alpha_1}(z) V_{\alpha_2}(w)
		\sim \sum_{\alpha_3 \in \cS_{\text{int}}} \frac{\widehat C_{\alpha_1, \alpha_2}^{\alpha_3}}{\abs{z - w}^{2 (\Delta_1 + \Delta_2 - \Delta_3)}} \sum_{N \in \N} \abs{z - w}^{2 N} \mc L_{-N} V_{\alpha_3}(w)\, ,
\end{equation} 
where the sum over $N$ generates the descendants of $V_{\alpha_3}$.
$\cS_{\text{int}}$ will be called the \textit{spectrum of internal states} or \textit{internal spectrum} for reasons that will become clear later, and includes all primary states which can appear in the sum for any $\alpha_1$ and $\alpha_2$.\footnotemark{}
\footnotetext{%
	Note that in general, $\alpha_1$ or $\alpha_2$ need not belong to $\cS_{\text{int}}$. We make a distinction between the external states and the internal states which will be important in our later discussion.
}%
Using the OPE, any $n$-point correlation function can be rewritten in terms of lower-order correlation functions.
Since the OPE coefficients $\widehat C_{\alpha_1, \alpha_2}^{\alpha_3}$ are related to the two-point and three-point functions, the decomposition stops once the $n$-point function is expressed in terms of the structure constants. 

An $n$-point correlation function of $n$ vertex operators on the sphere for the timelike theory has the form
\begin{equation}
	\label{eq:n-point}
	C_n(z_1, \alpha_1 ; \ldots ; z_n, \alpha_n)
		= \Mean{\prod_{i=1}^n V_{\alpha_i}(z_i)}\, .
\end{equation} 
Associativity of the OPE implies crossing symmetry constraints: an $n$-point function can be decomposed in many different channels, all of which should yield the same result.
In particular, the crossing symmetry constraint for the four-point function can be represented schematically as the equality between the `$s$-channel' and the `$t$-channel':
\begin{equation}
	\label{eq:crossing-4pt}
	\sum_{\alpha_s \in \cS_{\text{int}}}
			\vcenter{\hbox{\includegraphics{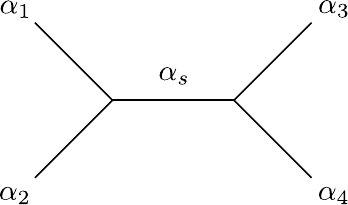}}}
	\qquad = \qquad
		\sum_{\alpha_t \in \cS_{\text{int}}}
			\vcenter{\hbox{\includegraphics{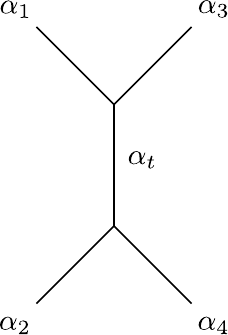}}}
\end{equation} 
These are a highly restrictive infinite set of integral constraints on the structure constants and on the allowed internal spectrum. 

The idea of the conformal bootstrap is to exploit these constraints to `solve' the theory by determining the structure constants completely with some additional physical restrictions on the spectrum as $(1)$ above, and requirements of analyticity or meromorphicity on the correlation functions as $(2)$. 
An impressive achievement of the conformal bootstrap has been to find a completely explicit analytic expression for the structure constants for both the spacelike and timelike regimes subject to these axiomatic requirements and show that they are unique. The spacelike structure constants $ C(a_{1} , a_{2}, a_{3})$ are given by the 
Dorn--Otto--Zamolodchikov--Zamolodchikov (DOZZ) formula~\cite{Dorn:1994:TwoThreepointFunctions, Zamolodchikov:1996:StructureConstantsConformal} (see~\cite{Teschner:1995:LiouvilleThreepointFunction, Teschner:2001:LiouvilleTheoryRevisited, Pakman:2006:LiouvilleTheoryAction, Ribault:2014:ConformalFieldTheory} for more discussions). The timelike structure constants $\widehat C(\a_{1} , \a_{2}, \a_{3})$ were obtained in~\cite{Schomerus:2003:RollingTachyonsLiouville, Zamolodchikov:2005:ThreepointFunctionMinimal, Kostov:2006:BulkCorrelationFunctions, Kostov:2007:NonRational2DQuantum-1, Kostov:2007:NonRational2DQuantum-2} (important insights were also provided in~\cite{Harlow:2011:AnalyticContinuationLiouville, Giribet:2012:TimelikeLiouvilleThreepoint}).
Since the correlation functions are meromorphic functions of the Liouville coupling, it is natural to analytically continue Liouville theory for any complex central charge $c_L \in \C$~\cite{Ribault:2015:LiouvilleTheoryCentral}.
It may appear then that it is possible to simply analytically continue spacelike Liouville into timelike Liouville~\cite{Strominger:2002:OpenStringCreation, Gutperle:2003:TimelikeBoundaryLiouville, Strominger:2003:CorrelatorsTimelikeBulk, Schomerus:2003:RollingTachyonsLiouville}. This however is not true~\cite{Fredenhagen:2003:MinisuperspaceModelsSbranes, McElgin:2008:NotesLiouvilleTheory, Zamolodchikov:2005:ThreepointFunctionMinimal} as we explain below.

A key ingredient in solving Liouville theory in the axiomatic framework are Teschner's \textit{difference} relations~\cite{Teschner:1995:LiouvilleThreepointFunction,Pakman:2006:LiouvilleTheoryAction}, also called \textit{degenerate} relations. These are a special case of the crossing relations where one of the external operators is a degenerate field. The general crossing symmetry constraints \eqref{eq:crossing-4pt}
form an infinite set of integral equations that are \textit{quadratic} in the structure constants. It is practically impossible to solve them in this form. 
However, these equations simplify enormously and become tractable if one of the external fields in the four-point function is taken to be a degenerate field of the Virasoro algebra.
The OPE with a degenerate field involves only a finite number of terms in the RHS of \eqref{eq:generic-ope} and as a result the crossing constraints 
reduce to effectively \textit{linear} shift relations. 
In particular, considering the two simplest degenerate fields leads to (schematically):
\begin{equation}
	\widehat C\left(\a_{1}, \a_{2}, \a_{3} \pm\b \right) = \widehat F(\b, \{\a_{i}\} ) \, \widehat C(\a_{1}, \a_{2}, \a_{3 }) \, ,
\end{equation} 
and a second identical equation but with $\b$ replaced by $1/\b$, which is a consequence of the $\b \rightarrow 1/\b$ symmetry of the theory. 
 $\widehat F(\b, \{\a_{i}\} )$ is a known function involving ratios of the Euler gamma function.
These degenerate crossing relations are already highly restrictive even though they are only a subset of the full set of crossing relations. In fact, they determine the structure constants completely for \textit{real} $\b$ by the following argument. A repeated application of Teschner relations results in a shift of $\a_{3}$ by a number $m\b/2 + n/2\b$ with $m, n \in \bZ$. This is dense in $\bR$ in that any real number can be approximated well enough by an appropriate choice of $m$ and $n$ because $1/\b$ and $\b$ are generically incommensurable. This makes it plausible that the structure constants can be uniquely determined for any value of $\a_{3} \in \bR$ by using these discrete difference relations. Then one can analytically continue to complex $\a_{3}$. 
A similar argument applies for the spacelike structure constants. 

If one analytically continues $\b$ (or $b$) to the complex plane away from the real line, then $\b$ and $1/\b$ are no longer collinear in the complex plane and the argument above does not work. In this case, the solution to the difference relations is ambiguous up to an elliptic function that is doubly periodic under shifts by $\b/2$ and $1/2\b$.
Consequently, the analytic continuations of $\widehat C(\a_{1} , \a_{2}, \a_{3})$ and $ C(a_{1} , a_{2}, a_{3})$ in the complex plane away from the real and imaginary axis can and do differ from each other by a doubly periodic function~\cite{Zamolodchikov:2005:ThreepointFunctionMinimal,Harlow:2011:AnalyticContinuationLiouville}. 
Schematically, one finds
\begin{equation}
\widehat C(a_{1} , a_{2}, a_{3}) = H(\{a_{i}\}, b) \, C(a_{1} , a_{2}, a_{3})
\end{equation}
where $H$ is a known elliptic meromorphic function which has poles for purely imaginary $b$ and hence one cannot reach it analytically from purely real $b$. Thus, even though one can analytically continue the Teschner relations, 
the structure constants $\widehat C(\a_{1} , \a_{2}, \a_{3})$ and $ C(a_{1} , a_{2}, a_{3})$ obtained as solutions of these equations are distinct and are not analytic continuations of each other.

Even though the \textit{degenerate} crossing relations determine the structure constants uniquely, to define a consistent theory it is necessary to show that even the \textit{full} crossing relations are satisfied for all choices of Liouville charges. This cannot be done analytically  except for $c_L \ge 25$~\cite{Teschner:2003:LectureLiouvilleVertex,Teschner:2001:LiouvilleTheoryRevisited} and $c_L = 1$~\cite{Gavrylenko:2018:CrossingInvariantCorrelation}, but in the impressive work of~\cite{Ribault:2015:LiouvilleTheoryCentral} Ribault and Santachiara developed a code to check it numerically for generic $c_L \in \C$. 

With these general considerations, in the next two subsections we review the spectrum and the correlation functions of timelike Liouville theory as determined from bootstrap.
We closely follow the conventions and presentation of~\cite{Ribault:2014:ConformalFieldTheory, Ribault:2015:LiouvilleTheoryCentral}. However, our perspective differs in two important respects.
\begin{itemize}
	\item
	First, as explained above, we regard the timelike and spacelike Liouville theories to be physically distinct,  defined respectively by the ranges $c_L \leq 1$ and $c_L \geq 25$ with real action. Thus our terminology is slightly different from~\cite{Ribault:2015:LiouvilleTheoryCentral}. 
	\item Second, and more importantly, we make a distinction between the spectrum of external and internal states as we explain in the next subsection. This point of view is more natural from the perspective of quantum gravity.
\end{itemize}

Before moving on, we add a comment on the bootstrap constraints at higher genus. On a higher-genus Riemann surface, inserting the resolution of the identity on a non-trivial cycle is equivalent to splitting the surface into two pieces with one additional state for each piece.
Demanding equality   between the original expression and the one with the identity inserted leads to what are known as the modular bootstrap constraints.
In general, a theory is consistent on any genus-$g$ Riemann surface if the four-point function on the sphere is crossing symmetric and if the one-point function on the torus is modular covariant for all states in the internal spectrum~\cite{Sonoda:1988:SewingConformalField-2}.
It has been shown that the Liouville theory~\cite{Hadasz:2010:ModularBootstrapLiouville, Ribault:2015:LiouvilleTheoryCentral} is a consistent theory even after taking into account these modular bootstrap constraints because the one-point function on the torus can be related to the four-point function on the sphere.

\subsection{On the Spectrum}
\label{sec:TLT:spectrum}
 
In this subsection,  we start by reviewing the usual parametrizations for the Liouville charges and the preliminary constraints on these obtained by simple considerations of reality and unitarity. We then present  a crucial distinction between the internal and external spectra that is an important and novel outcome of our analysis. 
Even though we are primarily interested in the timelike Liouville theory, we first discuss the spacelike Liouville theory because it is better known and also better understood from the perspectives of both the conformal bootstrap and minisuperspace quantization. 

The charges and the dimensions of the primary fields $\{ V_a \}$ of spacelike Liouville can be parametrized as
\begin{equation}
	\label{conformal-dimension-spacelike}
		\qquad
	a = \frac{Q}{2} + \I p,
	\qquad
	\Delta_a = a (Q - a) 	= \frac{Q^2}{4} + p^2 =: \Delta_p\, ,
\end{equation} 
where $a$ is the Liouville charge under the $\group{U}(1)$ current $J= \I \partial \phi$. Since the current is anomalous, 
the charge $a$ on the plane is related to the momentum $p$ in the cylinder as above.
We refer to $p$ as the Liouville momentum. When $Q$ is zero, it corresponds to the spacelike momentum in target space; in the minisuperspace quantization on the cylinder it corresponds to the asymptotic momentum away from the Liouville wall. 

\begin{figure}[h] 
	\begin{center}
	\vskip5mm
	\begin{subfigure}[c]{.45\linewidth}
		\centering
		\includegraphics[scale=0.4]{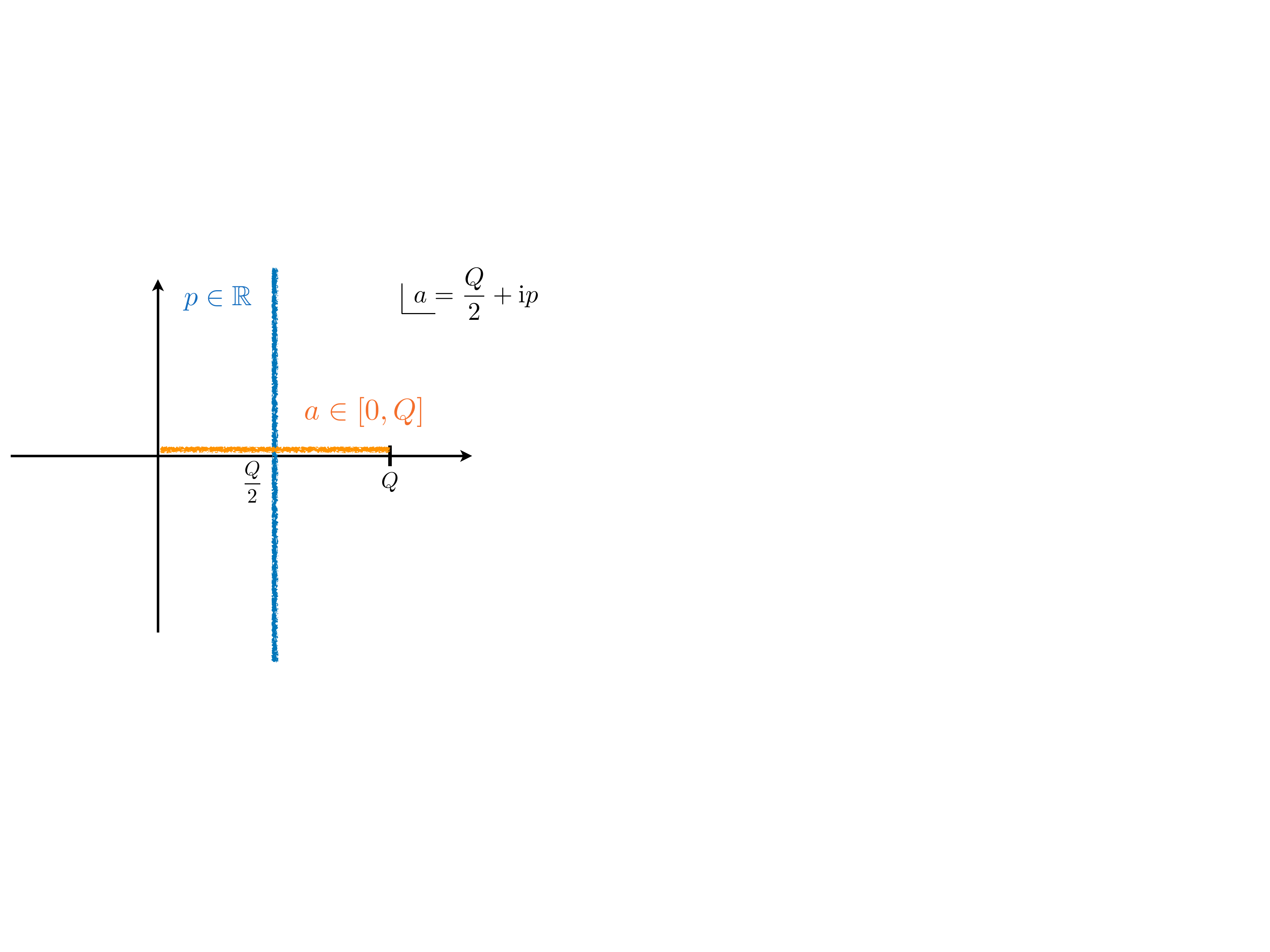}
		\subcaption{Charge plane}
	\end{subfigure}
	\begin{subfigure}[c]{.45\linewidth}
		\centering
		\includegraphics[scale=0.4]{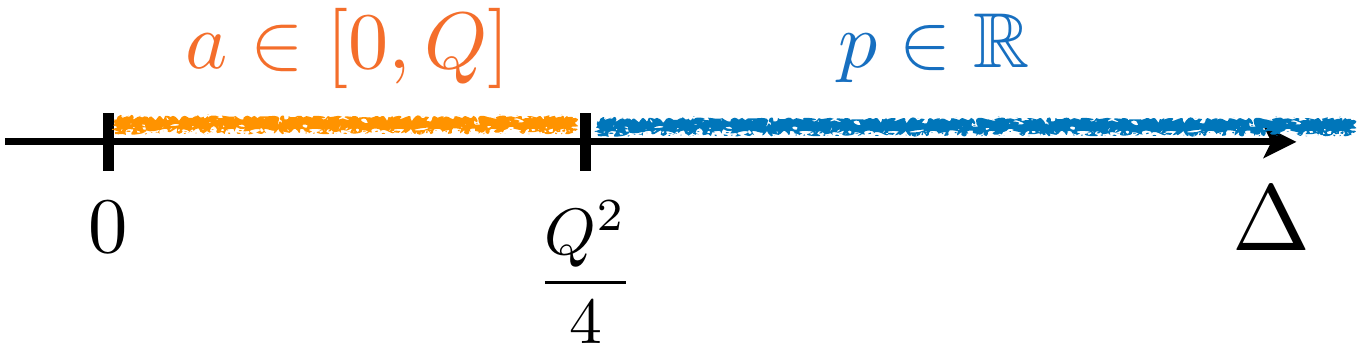}
		\subcaption{Conformal dimension plane}
	\end{subfigure}
	\caption{Two branches for the \emph{spacelike} Liouville charge $a$, and the corresponding conformal dimensions $\Delta_a$. Because of the reflection the ranges are halved to $a\in [0,Q/2]$ and $p\geq0$.
	}
	\label{spacelike spectrum}
	\end{center}
\end{figure}

To make contact with the Lagrangian approach, we note that the primary fields are in one-to-one correspondence with the exponential vertex operators with anomalous dimension given by the free Coulomb gas:
\begin{equation}
V_a = \e^{2 a \phi} \, ,
	\qquad
	\Delta_a= a (Q-a) \, .
	\label{exponential-spacelike}
\end{equation}
Since $c_L \geq 25$ for the spacelike regime, unitarity is a sensible assumption and consistency is confirmed by the bootstrap.
Unitarity implies that the conformal dimensions must be positive, which leads to two possible ranges for the spacelike momentum shown in Figure~\textbf{\ref{spacelike spectrum}}:
\begin{myenumerate}
\item
Real momentum $p \in \R $, shown in blue. 
\item
Imaginary momentum such that $a \in [0, Q]$, shown in orange.\footnote{For $c_L \geq 25$, the theory is unitary, which means the identity is the unique operator with $\Delta = 0$. Thus the operator with $a = 0$ must be identified with it. Even though it is not part of the internal spectrum, the associated state perfectly makes sense from the point of view of representation theory or as an external state. }
\end{myenumerate}

The expression for the conformal dimensions \eqref{conformal-dimension-spacelike} is invariant under $a\rightarrow Q-a$, or equivalently under $p \rightarrow -p$. 
Since the first axiom of Liouville theory about the spectrum requires that the spectrum be nondegenerate, the operators $V_a$ and $V_{Q-a}$ must be related. It is well known that the two can be identified inside correlation functions up to a `reflection coefficient' $R(a)$ by
\begin{equation}
	V_{a} = R(a) V_{Q-a} \, , \qquad R(a) R(Q-a) =1 \, .
\end{equation}
This implies that the range of $a$ has to be halved, restricting the orange branch to $a\in [0,Q/2]$ or $p\geq 0$ for the blue branch. 
In the minisuperspace quantization, the halving of the spectrum can be understood as a consequence of the exponential potential wall.
This potential wall reflects an `incoming' momentum $p$ into an `outgoing' momentum $-p$, so that a momentum state is only labelled by the norm of the momentum. With this identification, the two-point function can be completely diagonalized. 

For timelike Liouville, the charges and the dimensions of the primary fields $\{ V_\a \}$ can be parametrized as
\begin{equation}
	\alpha = - \frac{q}{2} + \I E \, ,
	\qquad
	\Delta_\alpha = \alpha (q + \alpha) = - \frac{q^2}{4} - E^2 =: \Delta_E \,. 
	\label{conformal-dimension-timelike}
\end{equation}
Throughout this paper, we use the terminology Liouville energy $E$ and Liouville momentum $p$ for timelike and spacelike Liouville respectively.
Now, the primary fields are in one-to-one correspondence with the exponential 
vertex operators with free Coulomb gas anomalous dimensions given by 
\begin{equation}
	V_\alpha = \e^{2\alpha \chi} \, ,
	\qquad
	\Delta_\alpha = \alpha (q + \alpha) \, ,\label{exponential-timelike}
\end{equation}
where $\alpha$ is the Liouville charge on the plane under the $\group{U}(1)$ current $J=- \I \partial \chi$, whose anomaly relates it to the Liouville energy $E$ on the cylinder as above. 

\begin{figure}[h] 
	\begin{center}
	\vskip5mm
	\begin{subfigure}[c]{.45\linewidth}
		\centering
		\includegraphics[scale=0.4]{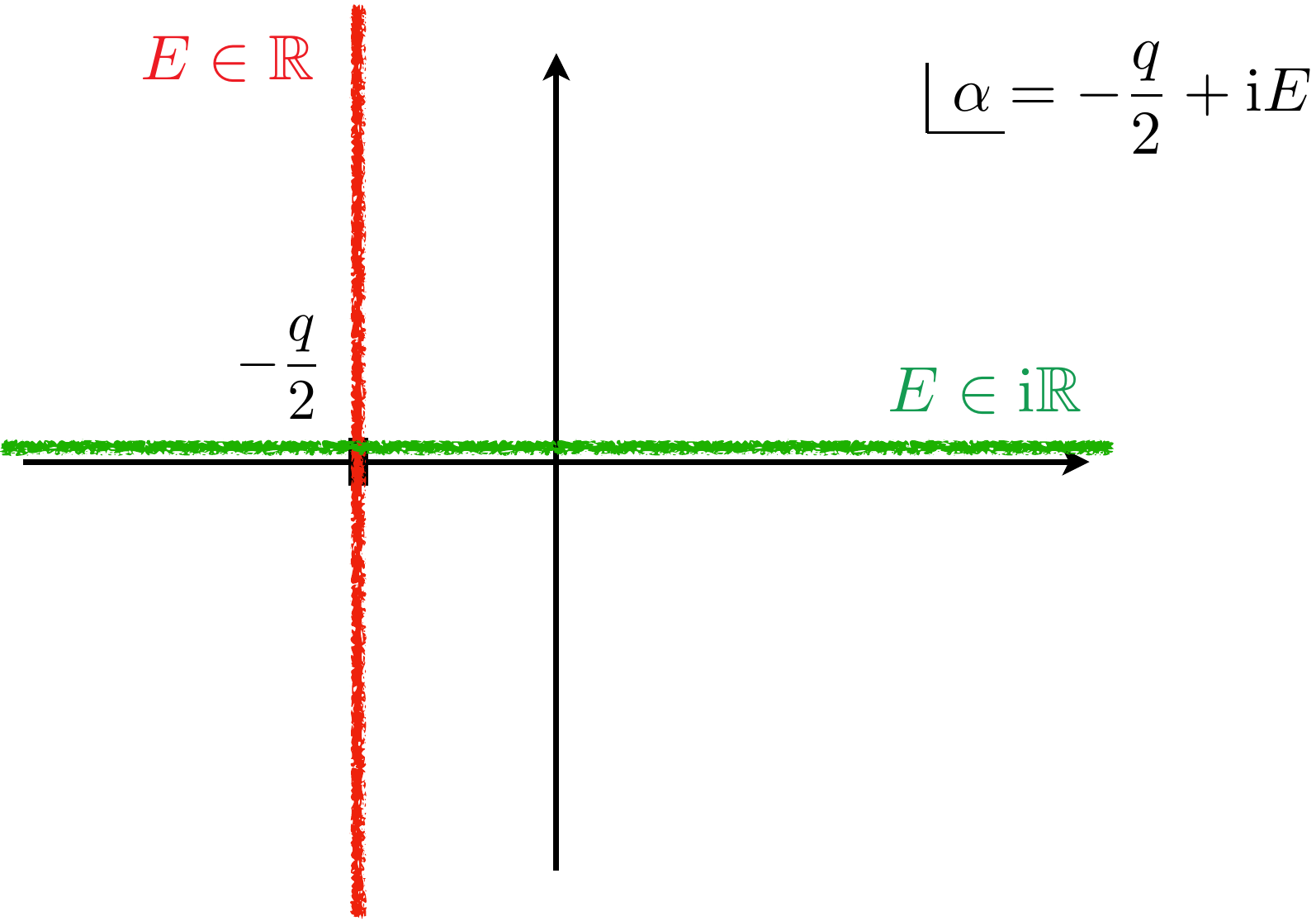}
		\subcaption{Charge plane}
	\end{subfigure}
	\begin{subfigure}[c]{.45\linewidth}
		\centering
		\includegraphics[scale=0.4]{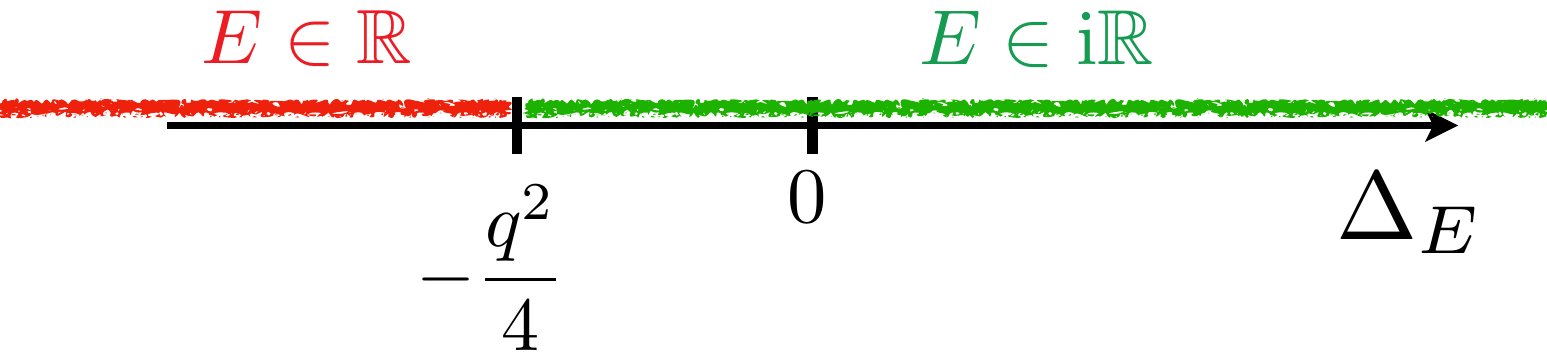}
		\subcaption{Conformal dimension plane}
	\end{subfigure}	
	\caption{Two branches of the \emph{timelike} Liouville charge $\alpha$ and the corresponding conformal dimensions $\Delta_E$. Because of the reflection, the ranges are halved to $E \geq 0 $ and $E \in \I [0, \infty)$.}
	\label{timelike spectrum}
	\end{center}
\end{figure}

For timelike Liouville, unitarity is no longer a sensible criterion and the conformal dimensions need no longer be positive. We only require that the conformal dimensions be real. This implies that either $E \in \R$ or $E \in \I \R$,
with the conformal dimensions bounded respectively from above or below by $- q^2/4$.
This leads to two possible ranges for the charge shown in 
Figure~\textbf{\ref{timelike spectrum}}:
\begin{myenumerate}
	\item
	Real energy $E \in \R $, shown in red. 
	\item
	Imaginary energy $E \in \I \R $, shown in green. 
\end{myenumerate}

The expression for the conformal dimensions \eqref{conformal-dimension-timelike} is now invariant under $\a\rightarrow -q -\a$, or equivalently under $E \rightarrow -E$. 
By the same reasoning as in the spacelike case, $V_\a$ and $V_{-q- \a}$ can be identified inside correlation functions up to a `reflection coefficient' $R(\a)$ by
\begin{equation}
	V_{\a} = R(\a) V_{-q-\a} \, , \qquad R(\a) R(-q-\a) =1 \, .
\end{equation}
This means that the range of $E$ is again halved, restricting to $E\geq 0$ for the red branch, or $E \in \I \bR^{+}$ for the green one. 

\vspace{5mm}

To define timelike gravity, we first need to determine the BRST cohomology.\footnotemark{}
\footnotetext{%
	The difficulties arising in the minisuperspace approximation are reviewed in \S\ref{sec:minisuperspace}.
}%
The resulting physical spectrum need not include all the states on the red and green branches as  allowed by the above considerations. Second, we  need to define the correlation functions for these physical states. These two questions naturally lead to an important distinction between external and internal states as we explain in the next subsections. We summarize here the main conclusions:
\begin{itemize}
\item \textit{External states} $\cS_{\text{ext}}$: These are the gauge-invariant states in the ghost-free BRST cohomology of the theory. We will find that their Liouville energies are a subset of $ \bR \cup \I \bR$.
\item \textit{Internal states} $\cS_{\text{int}}$: These are the states that appear in the internal energy integrals of correlators. They can be identified with the spectrum appearing in the OPE expansion, and are determined by requiring a convergent and crossing-symmetric four-point function. We will find that their Liouville energies belong to a subset of 
$ \I\R$. 
\end{itemize}

It may be surprising at first that $\cS_{\text{int}}$ does not include all the states that belong to $\cS_{\text{ext}}$. 
One would surely like to define correlation functions for all external states in the gauge-invariant spectrum of the theory. However, as we explain in §\ref{sec:TLT:corr}, the four-point function diverges when $E \in \R$ for the internal states,  as a consequence of  the behavior of the integrand for large negative conformal dimension ($E \to \pm \infty$).
One could try to consider only a finite interval, but there is no first principle to determine the bound. The only natural solution would have been the presence of a conservation law as for free scalars, but the $\mathrm{U}(1)$ invariance is broken by the background charge.
An unavoidable consequence of these facts is that $\cS_{\text{int}}$ cannot include states with real Liouville energies if we wish to obtain well-defined correlations. 
We adopt the point of view that the role of the bootstrap is to determine the minimal spectrum of internal states necessary to be able construct all physical correlation functions. It is possible, and is indeed true in the present case, that the correlation functions defined with a set of internal states make sense for a larger set of states.\footnotemark{}
\footnotetext{%
This type of reasoning can also be used to interpret the results from~\cite{Cao:2017:LiouvilleFieldTheory, Cao:2018:OperatorProductExpansion} for constructing the Gibbs measure of a Gaussian random field.
}%
In fact, the definition of the internal spectrum is somewhat fuzzy while considering a continuous spectrum since the integral over the states can be deformed as long as no singularity is encountered.
These two facts are of prime importance to define Liouville theory~\cite{Ribault:2014:ConformalFieldTheory, Ribault:2015:LiouvilleTheoryCentral} even at the level of the conformal field theory, as we clarify in the following two points.
\begin{itemize}
\item
First, to obtain the difference relations of the structure constants it is necessary to use degenerate external states, even though degenerate states do not belong to the  original set of internal or external states. We stress that a state $V_\alpha$ with $\Delta_\alpha$ equal to the dimension of a degenerate state is generically \textit{not} identified with the corresponding degenerate state.
The Verma module generated by $V_\alpha$ is in general a different representation than the one generated by the degenerate state.
Thus, using degenerate fields for deriving Teschner's relations amounts to extending both the internal  and external spectra to consider  more general OPE.
If the formulas  derived with this extended spectrum make sense for the original set of states,  for example, for the structure constants, then one can consider the theory only for the restricted set of states. This is the sense in which axiom 3.1 in~\cite{Ribault:2014:ConformalFieldTheory} should be interpreted.
\item
Second, even if one wants to identify the external spectrum with the internal one as in~\cite{Ribault:2015:LiouvilleTheoryCentral}, the internal spectrum \eqref{eq:ope-spectrum} needs to be slightly deformed in the complex plane anyway in order to get a convergent four-point function when $c_L \le 1$.
Any other analytic continuation of the contour yields the same results as long as no poles is encountered.
\end{itemize}

Even in quantum field theory or string field theory, the physical objects are S-matrix elements  as functions of  external momenta. The description in terms of Feynman diagrams and internal states is  convenient  but not essential. The internal states as such are not  observable. 
Asking if they are identical to the external states is not really a well-posed question.
From this perspective, there is no conceptual mystery if $\cS_{\text{int}}$ and $\cS_{\text{ext}}$ are distinct as long as  one can obtain sensible expressions for all physical quantities. 
Indeed, this is how things work in string field theory as we elaborate on in \S\ref{sec:sft}. The heuristic idea there is that as long as external  energies are finite, it is not necessary to analytically Wick rotate at infinity. It is sufficient to  analytically deform the contour  to avoid potential poles as external energies are analytically continued. One can thus prove Cutkosky rules  without explicitly talking about the internal states~\cite{Pius:2016:CutkoskyRulesSuperstring, Pius:2018:UnitarityBoxDiagram}.
We do not have the constraints from  Cutkosky rules in the present context but crossing symmetry is an analog which is similarly restrictive.

\subsection{Correlation functions}

We first review the two-point and the three-point functions. We then discuss the four-point functions with the view to determine the spectrum $\cS_{\text{int}}$ of internal states. 

\label{sec:TLT:corr}

\subsubsection*{Two-point function}

The vertex operators $\{ V_{\a} \}$ can be normalized so that 
the two-point function takes the form~\cite{Harlow:2011:AnalyticContinuationLiouville, Ribault:2014:ConformalFieldTheory}
\begin{equation}\label{2-point-fn}
	C_2(z_1, \alpha_1 ; z_2, \alpha_2)
		= \frac{1}{\abs{z_1 - z_2}^{4 \Delta_{\alpha_1}}} \,
			\big( \delta(q + \alpha_1 + \alpha_2) + R(\alpha_1) \, \delta(\alpha_1 - \alpha_2) \big) \, .
\end{equation} 
The dependence in $z_1$ and $z_2$ is completely fixed by global conformal invariance.
The presence of the reflection coefficient $R(\alpha)$ in \eqref{2-point-fn} is due to the reflection-invariance of the two-point function.
The reflection coefficient can be computed by considering the four-point function of two degenerate fields~\cite{Harlow:2011:AnalyticContinuationLiouville, Ribault:2014:ConformalFieldTheory} and reads
\begin{equation}
	R(\alpha)
		= - \left(\frac{ \e^{\I\pi }}{ -\pi \mu\, \gamma(-\beta^2)}\right)^{\frac{q+2\alpha}{\beta} }
			\frac{\Gamma\big(\beta (-q -2\alpha)\big) \Gamma\big(\beta^{-1} (q + 2\alpha)\big)}{\Gamma\big(\beta (q + 2\alpha)\big) \Gamma\big(\beta^{-1} (-q - 2\alpha)\big)} \, ,
\end{equation} 
with $\gamma(x) = \Gamma(x) / \Gamma(1-x)$.
The two-point function has a well-defined analytic continuation in $\beta$ and in $\alpha$. In particular, the normalization chosen here is an analytic continuation of the normalization for spacelike Liouville.

\subsubsection*{Three-point function}

Conformal invariance fixes the form of the three-point function to be
\begin{equation}
	C_3(z_1, \alpha_1 ; z_2, \alpha_2 ; z_3, \alpha_3)
		= \frac{\widehat C_{\alpha_1, \alpha_2, \alpha_3}}{\abs{z_1 - z_2}^{2 (\Delta_1 + \Delta_2 - \Delta_3)} \abs{z_2 - z_3}^{2 (\Delta_2 + \Delta_3 - \Delta_1)} \abs{z_3 - z_1}^{2 (\Delta_3 + \Delta_1 - \Delta_2)}}\, ,
\end{equation} 
where $\widehat C_{\alpha_1, \alpha_2, \alpha_3} := \widehat C(\alpha_1, \alpha_2, \alpha_3) $ are the structure constants of the theory. They are given by\footnotemark{}
\footnotetext{%
	The hat on the $\widehat C_{\alpha_1,\alpha_2, \alpha_3}$ is added to distinguish these structure constants in the timelike regime from the ones in the spacelike regime, for which $C_{a_1,a_2,a_3}$ is used in most of the literature.
	The hat reminds of the fact that these are two different functions of the Liouville momenta.
}%
\begin{equation}
	\label{eq:3pt-dozz}
	\widehat C_{\alpha_1, \alpha_2, \alpha_3}
		= \left(\frac{e^{\I\pi }}{ -\beta^{2 + 2\beta^2} \, \pi\mu \, \gamma(-\beta^2) }\right)^{\frac{q+\alpha}{\beta}} \,
			\frac{\Upsilon_\beta(\beta - q - \alpha)}{\Upsilon_\beta(\beta)} \,
			\prod_{i=1}^3 \frac{\Upsilon_\beta(\beta + 2\alpha_i - \alpha)}{\Upsilon_\beta(\beta - 2\alpha_i)} \, ,
\end{equation}
where $\alpha = \alpha_1 + \alpha_2 + \alpha_3$.
The Upsilon function $\Upsilon_{\beta}(x)$ is defined in §\ref{sec:cft:upsilon}.
This formula is valid for all $\alpha_i \in \C$.
This structure constant was found by Zamolodchikov, and independently by Kostov and Petkova~\cite{Zamolodchikov:2005:ThreepointFunctionMinimal, Kostov:2006:BulkCorrelationFunctions, Kostov:2007:NonRational2DQuantum-1, Kostov:2007:NonRational2DQuantum-2}. The expression at $c_L = 1$ already appeared in~\cite{Schomerus:2003:RollingTachyonsLiouville}.
More insights on this formula from the path integral perspective can be found in~\cite{Harlow:2011:AnalyticContinuationLiouville, Giribet:2012:TimelikeLiouvilleThreepoint}.

As already mentioned, the structure constant $\widehat C_{\alpha_1, \alpha_2, \alpha_3}$ is the unique solution to the degenerate crossing relations when $c_L \le 1$, and even if the degenerate relations admit a continuation to all $c_L \in \C$, $\widehat C$ can only be analytically continued to $c_L \notin (25, \infty)$. Given the zeros of the Upsilon function (see §\ref{sec:cft:upsilon}), the explicit form of $\widehat C_{\alpha_1, \alpha_2, \alpha_3}$ shows why analytic continuation to the spacelike regime $c_L \geq 25$ does not work, in line with the general remarks in \S\ref{Conformal}.

\subsubsection*{Four-point function}

Higher-point correlation functions can be constructed from the structure constants by using the OPE.
Concretely, the $s$-channel decomposition of the four-point function reads
\begin{equation}
	\label{eq:4pt}
	C_4(z_i, \alpha_i)
		= \int_{\cS_{\text{int}}} \dd \alpha_s \,
			\widehat C_{\alpha_1,\alpha_2, \alpha_s} \, \widehat C_{-q-\alpha_s,\alpha_3,\alpha_4} 
			\abs{\mc F^{(s)}_{\alpha_s}(z_i,\alpha_i)}^2
\end{equation}
where $\mc F^{(s)}_{\alpha_s}$ are the $s$-channel conformal blocks (see~\cite{Ribault:2014:ConformalFieldTheory} for a complete characterization).
The integral runs over the internal states, i.e.\ the internal spectrum (taken to be continuous by assumption).
The $t$- and $u$-channels are obtained similarly by considering different OPEs.

Requiring convergence of the decomposition \eqref{eq:4pt} imposes a restriction on the contour of integration, i.e.\ on the internal spectrum.
Indeed, the integrand behaves as $\abs{\mathfrak q}^{2 \Delta_s}$ with $\abs{\mathfrak q} < 1$ for large $\abs{\Delta_s}$ (see §\ref{sec:ana} for more details), hence the integral diverges when the (real part of the) conformal dimensions of the operators appearing in $\cS_{\text{int}}$ is unbounded from below.
For both the spacelike and timelike regimes, the respective conformal dimensions $\Delta_p=Q^2/4+p^2$ or $\Delta_E=-q^2/4-E^2$ are bounded from below when $E=\I p\in \I \R$.
Therefore, this continuous family of states can be identified to be the internal spectrum in both regimes (i.e.\ the family with $p\in \R$ in the spacelike case and the family with $E\in \I \R$ in the timelike case).
By continuity, this internal spectrum is also used for any $c_L \in \C$.
We stress that the internal spectrum is not analytically continued as one continues the field and the central charge. It yields sensible physical results even though it is na\"ively at odds with the conventional wisdom behind Wick rotation of the timelike mode in string theory.

\begin{figure}[ht]
	\centering
	\includegraphics[scale=1.5]{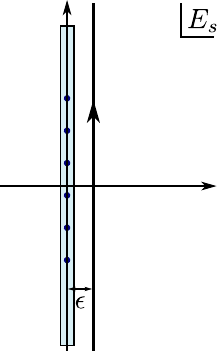}
	\caption{%
		Poles of the four-point function integrand and integration contour for the $c_L \le 1$ Liouville theory.
		The poles in the $E_s$-plane are located on the imaginary axis (shaded area, only few poles are displayed) and depend only on $\beta$, \emph{not} on the external momenta $\alpha_1, \alpha_2, \alpha_3$ and $\alpha_4$.
		The $\I\epsilon$ prescription shifts the contour away from the poles.
		}
	\label{fig:poles-4pt}
\end{figure}

However, there is a small caveat in the timelike regime: in this case, the so-identified internal spectrum includes states with dimensions equal to those of the degenerate states, which happen to correspond to the poles of the conformal blocks.
This is another indication that the case $c_L \le 1$ is subtler than the $c_L\ge 25$, and cannot be obtained by an analytic continuation from the latter.
The remedy to avoid the poles is to slightly shift the contour of integration by a small real number as~\cite{Ribault:2015:LiouvilleTheoryCentral, Ribault:2014:ConformalFieldTheory}
\begin{equation}
	\label{eq:ope-spectrum}
	\cS_{\text{int}}
		= \left\{ \alpha = - \frac{q}{2} + \I E, \, E \in \I \R + \epsilon \right\}.
\end{equation} 
The poles and integration contour are described in Figure~\textbf{\ref{fig:poles-4pt}}.
This prescription is equivalent to shifting the momentum on the cylinder $E$ by $\I \epsilon$, which can be interpreted as the standard $\I \epsilon$ prescription of QFT.
It is also consistent with the fact that a continuous internal spectrum can be deformed in the complex plane when no singularity is encountered.

The limit $\epsilon \to 0$ is taken in any expression which converges.
Moreover, since one can deform the contour of integration in the complex plane, the value of \eqref{eq:4pt} should be independent from $\epsilon$ as long as the expression converges and as long as the contour does not cross any poles.
Moreover, it is expected that the result does not depend on the sign of $\epsilon$ since the integrand is invariant under reflection $E_s\rightarrow -E_s$.
It has been checked numerically that this is indeed the case, and that the spectrum \eqref{eq:ope-spectrum} together with the structure constant $C$ and $\widehat C$ respectively for $c_L \notin (-\infty, 1]$ and $c_L \le 1$, lead to a crossing-symmetric four-point function for all $c_L \in \C$~\cite{Ribault:2015:LiouvilleTheoryCentral}.

\section{BRST cohomology and no-ghost theorem}
\label{sec:brst}

In this section, we consider the timelike Liouville theory coupled to $D \ge 25$ free scalar fields.
We closely follow~\cite{Bouwknegt:1992:BRSTAnalysisPhysical, Bilal:1992:RemarksBRSTcohomologycM, Polchinski:2005:StringTheory-1} and begin by replacing the Liouville theory by a Coulomb gas with charge $q$. In 
\S\ref{sec:brst:cft} we collect the relevant formulas for the Coulomb gas and ghost CFTs and then in \S\ref{sec:brst:cohom} compute the BRST cohomology.\footnotemark{}
\footnotetext{%
	Analysis of  BRST cohomology for more general matter systems will be presented  in~\cite{Bautista:BRSTCohomologyTimelike}.
}%
We focus on the holomorphic sector, and keep normal ordering implicit.
We comment on the effect of the cosmological constant in \S\ref{sec:brst:noghost} 
and prove the no-ghost theorem by imposing hermiticity of the matter sector. 

We follow the formalism of~\cite{Bouwknegt:1992:BRSTAnalysisPhysical} rather than the more recent and general method in~\cite{Asano:2000:NoghostTheoremString} which requires only a single flat timelike direction.  
Our reasons are two fold. 
First, the latter method does not  explicitly construct  the physical states~\cite{Asano:2000:NoghostTheoremString}.
Second, it relies on the “vanishing theorem”, which itself relies on the isomorphism between the Fock space and the Verma module of a scalar field; this isomorphism breaks down for $c < 1$ due to the existence of degenerate states.

\subsection{Coulomb gas and ghost CFTs}
\label{sec:brst:cft}

\subsubsection*{Coulomb gas}

The Coulomb gas action\footnote{We use the timelike kinetic term. Spacelike formulas follow by the analytic continuation \eqref{eq:analytic-cont-chi-q}.} for a scalar field $\chi$ with background charge $q$ is given by
\begin{equation}
	\label{Coulomb_gas_action}
	S = - \frac{1}{4\pi} \int \dd^2 x \sqrt{g} \,
		\left((\nabla \chi)^2 + q R \chi \right).
\end{equation}
The charge $q$ can be parametrized in terms of the parameter $\beta$ as \eqref{eq:q-beta}.
The energy--momentum tensor on a flat background and the central charge are
\begin{equation}
	T = (\pd \chi)^2 - q \, \pd^2 \chi,
	\qquad
	c_L = 1 - 6 q^2 \, .
\end{equation}
The action \eqref{Coulomb_gas_action} changes by a constant term under a constant shift of $\chi$. The corresponding conserved current $J(z) = - \I \pd \chi$ is anomalous at the quantum level if $q \neq 0$.

The Fourier expansion of the Coulomb gas field is
\begin{equation}\label{Fourier-expansion-Coulomb-field}
	\chi(z)
		= \frac{x}{2}
			+ \alpha \ln z
			+ \frac{\I}{\sqrt{2}} \sum_{n \neq 0} \frac{\alpha_n}{n} \, z^{-n}.
\end{equation}
The zero-mode $\alpha$ corresponds to the conserved charge of the current $J$.
Due to the anomaly of the current, this is related to the energy $E$ on the cylinder by
\begin{equation}
	\label{cft:eq:vertex-relation-a-p}
	\alpha = - \frac{q}{2} + \I E \, .
\end{equation}
The Virasoro operators are
\begin{equation}\label{Virasoro_Coulomb_gas}
	L_m
		= - \frac{1}{2} \sum_{n \neq 0} \alpha_n \alpha_{m-n}
			- \frac{\I}{\sqrt{2}} \big( \alpha + q (m + 1) \big) \alpha_m \, .
\end{equation}
The Virasoro zero-mode can be simplified to
\begin{equation}
	\label{Virasoro-zero-mode}
	L_0
		= N + \alpha (\alpha + q)
		= N - \frac{q^2}{4} - E^2,
\end{equation}
where $N$ is the level operator built from the number operators $N_n$ at each level $n > 0$
\begin{equation}
	N = \sum_{n > 0} n\, N_n \, ,\qquad N_n = - \frac{1}{n}\, \alpha_{-n} \alpha_n \,.
\end{equation}

The primary states of the theory are, as in timelike Liouville theory, the vertex operators $V_\alpha$
\begin{equation}
	\label{eq:primary-state}
	V_\alpha = \e^{2 \alpha \chi},
	\qquad
	\Delta_\alpha = \alpha \, (q + \alpha).
\end{equation} 
According to the anomalous shift \eqref{cft:eq:vertex-relation-a-p}, they correspond to the operators $V_E$ on the cylinder
\begin{equation}
	V_E = \e^{2 \I E \chi},
	\qquad
	\Delta_E = - \frac{q^2}{4} - E^2 .
\end{equation} 
The Fock vacua $\ket{E}$ are obtained by acting with the vertex operators on the $\group{SL}(2, \C)$ vacuum $\ket{0}$
\begin{equation}
	\ket{E} = V_E \ket{0}.
\end{equation}
The Fock space is generated by applying creation operators $\alpha_{-n}$ with $n > 0$ on the vacuum
\begin{equation}
	\ket{\psi} = \prod_{n \ge 1} (\alpha_{-n})^{N_n} \ket{E},
	\qquad
	N_n \in \N.
\end{equation}

\subsubsection*{Ghosts}

The action is \eqref{eq:action-ghosts}.
The energy--momentum tensor, central charge and conformal dimensions of the ghosts read
\begin{equation}
	T^{\text{gh}}
		= - \pd(b c) - b \pd c,
	\qquad
	c_{\text{gh}} = - 26,
	\qquad
	\Delta_b = 2,
	\qquad
	\Delta_c = - 1.
\end{equation}
The ghost action is invariant under an anomalous $\group{U}(1)$ global symmetry with current $j(z) = - b\, c$ and the associated charge is called the ghost number $N^{\text{gh}}$, normalised such that
\begin{equation}
	N^{\text{gh}}(b) = - 1, \qquad
	N^{\text{gh}}(c) = 1
\end{equation}
on the plane.

The mode expansions are
\begin{equation}
	b(z) = \sum_n b_n\, z^{- n - 2}, \qquad
	c(z) = \sum_n c_n\, z^{- n + 1},
\end{equation}
in terms of which the Virasoro modes are
\begin{equation}
	L^{\text{gh}}_m = \sum_n (m - n)\, b_{m+n} c_{-n} \, ,
	\qquad
	L^{\text{gh}}_0 = N^b + N^c - 1 \, ,
\end{equation}
where the level operators $N^b$ and $N^c$ are built from the ghost number operators at level $n > 0$
\begin{equation}
	N^b=\sum_{n>0} n\, N_n^b\, ,\quad N_n^b = b_{-n} c_n \, ,
	\qquad
	N^c=\sum_{n>0} n \,N_n^c\, ,\quad N_n^c = c_{-n} b_n \, .
\end{equation}

The $\group{SL}(2, \C)$-invariant vacuum $\ket{0}$ is defined by
\begin{equation}
		b_n \ket{0} = 0 \, , \quad \forall \,  n \ge -1 \, ; \qquad
		c_n \ket{0} = 0\, , \quad 	
	\forall \,  n \ge 2 \, .
\end{equation}
However, there exists a $2$-fold-degenerate state with a lower energy since the $\group{SL}(2, \C)$ vacuum is not annihilated by $c_1$, which is a positive-frequency mode.
The degeneracy arises because $b_0$ and $c_0$ commute with the Hamiltonian.
The two ground states are given by
\begin{equation}
	\ket{\downarrow} = c_1 \ket{0}, \qquad
	\ket{\uparrow} = c_0 c_1 \ket{0}.
\end{equation}
By analogy with the critical string, we take $\ket{\downarrow}$ to build the Fock space by acting on it with the creation operators
\begin{equation}
	\ket{\psi}
		= c_0^{N_0^c} \prod_{n \ge 1} (b_{-n})^{N_n^b} (c_{-n})^{N_n^c}\, \ket{\downarrow},
	\qquad
	N_n^b, N_n^c = 0, 1\,.
\end{equation}

\subsection{BRST cohomology}
\label{sec:brst:cohom}

The main idea behind the analysis of the BRST cohomology is as follows. 
Instead of working out the cohomology of the operator $Q_{B}$, it is easier to work out the cohomology of simpler operators $\widehat Q$ and $Q_{0}$ which we introduce below. If the cohomology of  $Q_{0}$ is ghost-free 
(as we demonstrate in the next subsection under suitable conditions), then it is isomorphic to the cohomology of $\widehat Q$ and in turn to the  `relative' cohomology of  the original BRST operator $Q_{B}$.  The `absolute' cohomology of $Q_{B}$ involves a trivial doubling of the relative cohomology as we explain below.

We consider a CFT constructed from three sectors: the timelike Liouville field $\chi$, $D$ spacelike free scalar fields $(X, Y^i)$ with $i = 1, \ldots, D - 1$, and the $b$ and $c$ ghost fields. The separation of one scalar $X$ from the other $Y^i$ is necessary to define light-cone coordinates in field space.
We call longitudinal $\|$ the sector made from the $(\chi, X, b, c)$ fields, and transverse $\perp$ that of the $\vec Y$ fields. 

The mode operators of the field $\chi$ are denoted by $\alpha^\chi_n$, and its zero mode $\alpha^\chi$ is related to its energy on the cylinder $E$ by \eqref{cft:eq:vertex-relation-a-p}. For the spacelike scalar $X$, the mode expansion is analogous to that of the Coulomb gas \eqref{Fourier-expansion-Coulomb-field}, but consistently with our spacelike notation we denote the modes by $a^X_n$, and the zero-mode by $a^X=\I K$, where $K$ is its momentum on the cylinder. For the fields $\vec Y$, their momenta on the cylinder are denoted by $\vec k$, and we don't need to introduce their Fourier modes.

A general state of the longitudinal sector is
\begin{equation}
	\ket{\psi_\|} = c_0^{N_0^c} \prod_{m > 0} (\alpha^\chi_{-m})^{N_m^\chi} (a^X_{-m})^{N_m^X}\, (b_{-m})^{N_m^b} (c_{-m})^{N_m^c}\, \ket{E, K, \downarrow}
\end{equation}
with $\ket{E, K, \downarrow} := \ket{E} \otimes \ket{K} \otimes \ket{\downarrow}$.

The total Virasoro zero-mode is
\begin{equation}
	\label{eq:L0-total}
	L_0 = \vec k^2 + K^2 - E^2 - \frac{q^2}{4} - 1
		+ \what L_0^\| + \what L_0^\perp \, ,
\end{equation}
where $\what L_0^\|, \what L_0^\perp \in \N$ are the total level operators
\begin{equation}
	\label{eq:level-operator}
	\what L_0^\| = N^\chi + N^X + N^b + N^c, \qquad
	\what L_0^\perp = \sum_{i=1}^{D-1} N^i,
\end{equation}
with $N^i$ counting the $Y^i$ modes.
The vanishing of the total central charge allows to express $q$ in terms of $D$ 
\begin{equation}
	q^2 = \frac{D - 25}{6} \, .
\end{equation}

The mode expansion of the BRST charge $Q_B$ is
\begin{equation}
	\label{general_BRST_charge}
	Q_B
		= \sum_n c_{n} L^m_{-n} + \frac{1}{2} \sum_{m,n} (n - m)\, c_{-m} c_{-n} b_{m+n} - c_0,
\end{equation} 
where $L^m_n$ are the total matter\footnote{By matter here we mean all the fields except for the ghosts, i.e.\ including the Liouville field, since the CFT doesn't `\textit{know}' that the Liouville field comes from the gravity sector.} Virasoro modes. 
The BRST charge is nilpotent $Q_B^2 = 0$ and has ghost number $N_{\text{gh}}(Q_B) = 1$.
Importantly, the total Virasoro operators are exact and commute with the BRST charge
\begin{equation}
	\label{Virasoro_BRST_ghost}
	L_n = \anticom{Q_B}{b_n},
	\qquad
	\com{Q_B}{L_n} = 0.
\end{equation}

Physical states are defined to be those in the (absolute) BRST cohomology $\mc H(Q_B)$ of $Q_B$, i.e.\ closed but non-exact states.
A necessary condition for a state $\ket{\psi}$ to be in the BRST cohomology is to be on-shell, i.e.\ its conformal dimension $\Delta$ has to vanish: 
\begin{equation}
	\label{eq:on-shell}
	L_0 \ket{\psi} = 0.
\end{equation}
Indeed, suppose that $\ket{\psi}$ is closed but not on-shell, then one can use \eqref{Virasoro_BRST_ghost} to write
\begin{equation}
	\label{eq:off-shell-psi}
	\ket{\psi}
		= \frac{1}{L_0}\, Q_B \big( b_0 \ket{\psi} \big).
\end{equation}
The state $b_0 \ket{\psi}$ corresponds to another state in the Fock space and thus $\ket{\psi}$ is exact and does not belong to the cohomology.

A stronger condition that can be imposed on a closed state $\ket{\psi}$ to avoid such exact states is $b_0 \ket{\psi} = 0$.
Thanks to \eqref{Virasoro_BRST_ghost}, this implies $L_0 \ket{\psi}= 0$ (but the converse is not true). Imposing this condition is useful for the following reason.
The BRST charge can be decomposed into the ghost zero-modes as
\begin{equation}
	\label{eq:splitting-Q}
	Q_B = c_0 L_0 - b_0 M + \what Q \, ,
\end{equation}
where
\begin{subequations}
\begin{gather}
	\label{eq:Q-hat}
	\what Q = \sum_{n \neq 0} c_{-n} L^m_n
		- \frac{1}{2} \sum_{\substack{m,n \neq 0 \\ m + n \neq 0}} (m - n)\, c_{-m} c_{-n} b_{m+n}\ , \\
	M = \sum_{n \neq 0} n\, c_{-n} c_n \, .
\end{gather}
\end{subequations}
In the subspace $b_0 = 0$, the BRST charge reduces to $\what Q$, which in turn is nilpotent in this subspace.
Hence, finding the cohomology $\mc H(Q_B) \cap \ker b_0$ is equivalent to computing the cohomology $\mc H(\what Q)$, so-called relative cohomology.
It is well known that the absolute cohomology is simply $\mc H(Q_B) = \mc H(\what Q) \oplus c_0 \mc H(\what Q)$~\cite{Bouwknegt:1992:BRSTAnalysisPhysical}.

In order to simplify further the problem, we introduce the following light-cone parametrization of the longitudinal matter:
\begin{equation}
	\begin{gathered}
	\alpha^\pm_n
		= \frac{1}{\sqrt{2}}\, (\alpha^\chi_n \pm a^X_n),
	\qquad
	\alpha^\pm
		= \frac{1}{\sqrt{2}}\, (- \alpha^\chi \pm a^X),
	\\
	x^\pm
		= \frac{1}{\sqrt{2}}\, ( - x^\chi \pm x^X), \qquad
	K^\pm
		= \frac{1}{\sqrt{2}}\, (- E \pm K) \, .
	\end{gathered}
\end{equation}
Using the relation between the zero modes and the energy and momentum, i.e.\ $a^X=\I K$ and \eqref{cft:eq:vertex-relation-a-p}, one obtains
\begin{equation}
	\alpha^\pm = \frac{q}{2 \sqrt{2}} + \I K^\pm.
\end{equation}

The expression \eqref{eq:L0-total} for the total Virasoro zero-mode becomes
\begin{subequations}
\begin{gather}
	\label{eq:on-shell-condition}
	L_0 = \vec k^2 - \frac{K^+ K^-}{2} - \frac{q^2}{4} - 1 
		+ \what L_0^\| + \what L_0^\perp, \\
	\label{eq:level-operator-lc}
	\what L_0^\| = N^+ + N^- + N^b + N^c \, ,
	\end{gather}
\end{subequations}
where we defined the light-cone level and number operators as
\begin{equation}
	N^\pm = \sum_{n > 0} n\, N^\pm_n, \qquad
	N^\pm_n = - \frac{1}{n}\, \alpha_{-n}^\pm \alpha_n^\mp.
\end{equation}
We also define generalized momenta as
\begin{equation}
	\label{eq:P-pm-n}
	K^\pm_m = - \I \alpha^\pm + \frac{\I q}{2 \sqrt{2}} (m + 1)
		= K^\pm + \frac{\I q \,m}{2 \sqrt{2}}\,,
\end{equation}
such that $K^\pm_0 = K^\pm$.

In this light-cone parametrization, we decompose $\what Q$ as
\begin{equation}
	\label{eq:splitting-hat-Q}
	\what Q = Q_0 + Q_1 + Q_2
\end{equation}
where
\begin{equation}
	\begin{gathered}
		Q_1 = \sum_{m \neq 0} c_{-m} L^\perp_m
			+ \sum_{\substack{m,n \neq 0 \\ m + n \neq 0}} c_{-m} \left(- \alpha^+_{-n} \alpha^-_{m+n} - \frac{1}{2}\, (m - n)\, c_{-n} b_{m+n} \right) \, ,
		\\
		Q_0 = - \sqrt{2} \sum_{m \neq 0} K^+_m\, c_{-m} \alpha^-_m \, ,
		\qquad
		Q_2 = - \sqrt{2} \sum_{m \neq 0} K^-_m\, c_{-m} \alpha^+_m \, .
	\end{gathered}
\end{equation}
The interest of this decomposition is that nilpotency of $\what Q$ implies that $Q_0$ and $Q_2$ are also nilpotent, so they both define a cohomology.
One can show that the cohomologies $\mc H(\what Q)$ and $\mc H(Q_0)$\footnotemark{} are isomorphic under general conditions, which hold, in particular, when there are no ghosts~\cite{Bouwknegt:1992:BRSTAnalysisPhysical}.
\footnotetext{%
	An  equivalent statement holds with  $Q_0$ replaced by $Q_2$ by exchanging $K^\pm_n$.
}%

To determine the cohomology, we need to invert the generalized momentum $K^+_n$. For this reason, we will deal separately with the two cases where $K^+_n \neq 0$ for all $n$ ($K^-_n$ can vanish for some value) and where $K^+_r = K^-_s = 0$ for some $r, s \neq 0$.
The case where $K^-_n$ never vanishes is similar to the first one by reversing the definition of the degree. In the case where $K^+_r = K^-_s = 0$   there is a     subtlety because  the vanishing of the momenta implies that some oscillators are absent from the BRST operators $Q_0$ and $Q_2$.

\subsubsection*{Non-vanishing $K^\pm_n$: standard states}

If $K^+_n \neq 0$ for all $n$, one can introduce the operator
\begin{equation}
	B = \frac{1}{\sqrt{2}} \sum_{n \neq 0} \frac{1}{K^+_n} \, \alpha^+_{-n} b_n
\end{equation}
such that
\begin{equation}
	\what L_0^\| = \anticom{Q_0}{B}.
\end{equation}
A state $\ket{\psi}$ can be in the BRST cohomology only if
\begin{equation}
	\label{on-shell_reduced}
	\what L_0^\| \ket{\psi} = 0.
\end{equation}
Indeed, following the same reasoning as in \eqref{eq:off-shell-psi}, one finds that $\ket{\psi}$ is $Q_0$-exact
\begin{equation}
	\ket{\psi} = \frac{1}{\what L_0^\|} \, Q_0 \big( B \ket{\psi} \big)
\end{equation}
if $\what L_0^\| \ket{\psi} \neq 0$.

Since $\what L_0^\|$ is a sum of positive integers \eqref{eq:level-operator}, each term must be independently zero.
This implies that the states $\ket{\psi} \in \mc H(Q_0)$ cannot contain any longitudinal excitation
\begin{equation}
	N^\chi = N^X = N^b = N^c = 0,
\end{equation} 
and correspond to the ground state of the Fock space $\ket{E, K, \downarrow}$. 
Furthermore, states created with transverse excitations alone cannot be exact. Therefore,  the above condition $\what L_0^\| = 0$ is sufficient for these states to be nonexact.

The next step is to prove that $\what L_0^\| = 0$ states are $Q_0$-closed, such that this condition is sufficient also for the cohomology. 
These states have $N_{\text{gh}} = 1$ since they contain only the ghost vacuum $\ket{\downarrow}$.
Since $\what L_0^\|$ and $Q_0$ commute, one has
\begin{equation}
	0 = Q_0 \what L_0^\| \ket{\psi}
		= \what L_0^\| Q_0 \ket{\psi}.
\end{equation}
Since $Q_0$ increases the ghost number of $\ket{\psi}$ by $1$, one can invert $\what L_0^\|$ in the last term since $\what L_0^\| \neq 0$ in this subspace.
This gives $Q_0 \ket{\psi} = 0$, as wanted.

It remains to solve the on-shell condition \eqref{eq:L0-total}, which for these states reads
\begin{equation}
	\label{eq:on-shell-standard-states}
	L_0
		= \vec k^2 + K^2 - E^2 - \frac{q^2}{4} - 1
			+ \sum_{i=1}^{D-1} N^i
		= 0.
\end{equation}
This equation admits the solutions $E \in \R$, but also some $E \in \I \R$ thanks to the factor $-q^2$, namely
\begin{equation}
E\in  \left[- \frac{\I}{2}\sqrt{q^2+4},\frac{\I}{2}\sqrt{q^2+4}\right] \, ,
\end{equation}
and the bounds are saturated when the matter is in its ground state $N^i = 0$.
In particular, in the semiclassical limit $q \to \infty$, this includes all $E\in\I\R$.
We refer to the corresponding states as \textit{standard states}. They do not have ghosts and have positive norm, since they only have perpendicular excitations.

\subsubsection*{Vanishing $K^\pm_n$: discrete states}

If there exist two non-zero integers $r$ and $s$ such that the operators $K^\pm_n$ vanish
\begin{equation}
	\label{eq:vanishing-Pr-Ps}
	\exists\, r, s \in \Z^*:
		\qquad
		K^+_r = K^-_s
			= 0,
\end{equation}
one must introduce the modified operators
\begin{equation}
	B_r = \frac{1}{\sqrt{2}} \sum_{n \neq 0, r} \frac{1}{K^+_n}\, \alpha^+_{-n} b_n \, ,
	\qquad
	\what L_{0,r}^\| = \anticom{Q_0}{B_r}.
\end{equation}
By the same argument as in the previous case, a state $\ket{\psi}$ is in the cohomology only if
\begin{equation}
	\what L_{0,r}^\| \ket{\psi} = 0.
\end{equation}

We distinguish two cases, whether $r < 0$ or $r > 0$.
In each case, the zero-mode Virasoro and the reduced one are related as
\begin{equation}
	\label{eq:discrete-states-hat-L}
	\what L_0^\| =
		\begin{cases}
			\what L_{0,r}^\|+r\, \big( N^+_r + N^c_r \big) & r > 0 \\
			\what L_{0,r}^\| -r\, \big( N^-_{-r} + N^b_{-r} \big) & r < 0
		\end{cases}
\end{equation}
Since a state in the cohomology only requires $\what L_{0,r}^\|\ket{\psi}=0$, it can be built by acting with the corresponding creation operators on the vacuum:
\begin{itemize}
	\item $r > 0$ : $b_r$ is an annihilation operator associated to $c_{-r}$ and $\alpha^+_{-r}$ is a creation operator
	\begin{equation}\label{discrete-states-1}
		\ket{\psi}=(\alpha^+_{-r})^\mu\, (c_{-r})^\nu \ket{E, K, \downarrow}
		\end{equation}
	\item $r < 0$ : $b_r$ is a creation operator and $\alpha^+_{-r}$ is an annihilation operator associated to $\alpha^-_r$
	\begin{equation}\label{discrete-states-2}
		\ket{\psi}=(\alpha^-_r)^\mu\, (b_r)^\nu \ket{E, K, \downarrow}
		\end{equation}
\end{itemize}
where $\mu \in \N$ and $\nu = 0, 1$ are some non-negative integers to be determined in each case by consistency with other conditions.

Using the expression for the generalized momenta $K^\pm_m$ \eqref{eq:P-pm-n} (or more simply subtracting $K^+_r$ and $K^-_s$ directly, noting that every term which does not multiply $m$ disappears) one finds the simpler expressions 
\begin{equation}
	K^+_m = \frac{\I q}{2 \sqrt{2}} \, (m - r), \qquad
	K^-_m = \frac{\I q}{2 \sqrt{2}} \, (m - s)
\end{equation}
which lead to
\begin{equation}
	\label{eq:momentum-discrete-states}
	E = \frac{\I q}{4} (r + s) \, ,
	\qquad
	K = - \frac{\I q}{4} (r - s)
\end{equation}
and
\begin{equation}
	K^+ K^- = - \frac{r s}{2} \, \frac{D - 25}{24} .
\end{equation}
The on-shell condition on these states then yields
\begin{equation}
	\label{eq:on-shell-discrete-states}
	L_0
		= \vec k^2 - (1 - r s) \, \frac{D - 25}{24}  - 1 + \abs{r} (\mu + \nu)  + \sum_{i=1}^{D-1} N^i
		= 0.
\end{equation}
The cohomology will hence contain states with ghost number $N_{\text{gh}} = 1 + (\sign r) \nu = 0, 1, 2$.

\subsection{No-ghost theorem}
\label{sec:brst:noghost}

The standard states are ghost-free but the discrete states \eqref{discrete-states-1} and \eqref{discrete-states-2} can lead to negative-norm states. There are two types of discrete states which we treat separately.
\begin{itemize}
\item
States with $r \neq s$:

The discrete states with $r \neq s$ are removed by using the unitarity of the matter sector. The matter scalar fields $X$ and $\vec Y$ are hermitian, and their momenta are real, $K, \vec k \in \R$.
According to \eqref{eq:momentum-discrete-states}, for the discrete states with $r \neq s$, the momentum of $X$ would have to be imaginary, $K \in \I \R$ because $q \in \R$. 
Hence, these states are not allowed by hermiticity.
The matter fields $X$ and $\vec Y$ are unitary with positive inner-product. Hence, there is no negative-norm states once the timelike and ghost oscillators are removed.

\item
States with $r \neq -s$:

The discrete states with $r = s$ (and in fact all states $r \neq - s$) of the Coulomb gas theory are actually not present in the Liouville theory once the effect of the cosmological constant is taken into account. The cosmological constant wall breaks the shift symmetry. 
As explained in section \S\ref{sec:TLT}, this is taken into account by the reflection property which halves the energy spectrum.
Practically this means that rather than the set of Fock vacuum states $\ket{E}$ of the Coulomb gas, one should consider the linear combinations
\begin{equation}
	\ket{\mathcal{E}} = \ket{E} + R(E) \ket{-E},
\end{equation}
where $R(E)$ is the reflection coefficient. 
The states $\ket{\mathcal{E}}$ are no longer eigenstates of the `momentum operator in field space' or the shift operator $-\I\ \dd/\dd\chi_{0}$ except if $E = 0$.
But this implies $r = - s$ according to \eqref{eq:momentum-discrete-states}, and thus excludes the states $r = s$. 
\end{itemize}

Note that the states $\ket{\mathcal{E}}$ continue to be the eigenstates of the `Hamiltonian operator in the field space' which in the minisuperspace approximation involves a second derivative $- \dd^{2}/\dd\chi_{0}^{2}$ with respect to the zero mode $\chi_{0}$ of the Liouville field. 
The Virasoro generator $L_0$ contains only the `Hamiltonian of the zero mode' and hence is quadratic in the energy:
\begin{equation}
	L_0^\chi \ket{\mathcal{E}}
		= - \left( \frac{q^2}{4} + E^2\right) \ket{\mathcal{E}} 
\end{equation}
and allows for both $E \in \R \cup \I \R$ when $q \in \R$.
The derivation of the standard states which relied only on the $L_0$ eigenstates, 
is thus unaffected by the halving of the Fock space. 

In conclusion, with the inclusion of the cosmological term and by demanding unitarity of the matter sector, only the standard states are allowed and the spectrum is ghost free. The on-shell equation \eqref{eq:on-shell-standard-states} allows for  $E \in \R$ and a subset of $E \in \I \R$ as solutions, which we identify as the spectrum of external states:
\begin{equation}\label{eq:ext-spectrum}
\cS_{\text{ext}} = \left\{ \alpha = - \frac{q}{2} + \I E, \, E\in \R \,\cup \,\, \left[- \frac{\I}{2}\sqrt{q^2+4},\frac{\I}{2}\sqrt{q^2+4}\right]\right\}\, .
\end{equation}
In the semiclassical limit $q\rightarrow \infty$, $\cS_{\text{ext}} = \left\{ \alpha = - \frac{q}{2} + \I E, \, E\in \R \,\cup \,\,\I \R\right\}$.
Another consequence is that physical states do not include descendants of the Liouville primary operators $V_\alpha$ since the longitudinal sector cannot be excited.\footnote{However, they would appear in intermediate channels since they are built from the conformal blocks.
	This is a second aspect in which the internal and external states are different.} 
This proves the no-ghost theorem for all physical states in the BRST cohomology.

\section{Analytic continuation}
\label{sec:ana}

We now turn to the question of whether the four-point function defined using $\cS_{\text{int}}$ \eqref{eq:ope-spectrum} for internal energy integrals, can be analytically continued for all external states in $\cS_{\text{ext}}$ \eqref{eq:ext-spectrum}. For this purpose it is necessary to examine the asymptotic divergences and poles of the four-point function integrand
with  the two- and three-point functions  defined by the analytic continuation of  formulas as in \S\ref{sec:TLT:corr}.

\subsection{Divergence in the four-point function}

The large conformal dimension behaviour of the conformal block is~\cite{Zamolodchikov:1987:ConformalSymmetryTwodimensional, Zamolodchikov:1996:StructureConstantsConformal, Harlow:2011:AnalyticContinuationLiouville, Ribault:2015:LiouvilleTheoryCentral}
\begin{equation}
	\label{eq:cblock-large-p}
	\mc F^{(s)}_{\alpha_s}(z_i,\alpha_i)
		\sim_{\abs{\Delta_{\alpha_s}} \to \infty} (16 \mathfrak q)^{\Delta_{\alpha_s}}
\end{equation} 
where the elliptic nome $\mathfrak q$ is defined by
\begin{equation}
	\mathfrak q(x) = \exp\left( - \pi \, \frac{K'(x)}{K(x)} \right)
\end{equation} 
in terms of the complete elliptic integral of the first kind $K(x)$.
One has $\abs{\mathfrak q} < 1$ for all $x \in \C$.
Writing $\alpha_s = - \frac{q}{2} + \I E_s$ and combining together \eqref{eq:cblock-large-p} and \eqref{eq:upsilon-large-p}, the large $E_s$ behaviour of the integrand of the four-point function \eqref{eq:4pt} is found to be
\begin{equation}
	\label{eq:4-pt-divergence}
	\widehat C_{\alpha_1, \alpha_2, - q - \alpha_s} \, \widehat C_{\alpha_s, \alpha_3, \alpha_4} \abs{\mc F^{(s)}_{\alpha_s}(z_i,\alpha_i)}^2
		\sim_{\abs{E_s} \to \infty} \abs{\mathfrak q}^{- 2 E_s^2}
		\propto \abs{\mathfrak q}^{L_0 + \bar L_0}.
\end{equation} 
This behavior implies that the four-point function is finite for $\alpha_s \in \R$ or equivalently $E_s \in \I \R$ up to the $\I\epsilon$ prescription. It diverges for $\alpha_s \in - \frac{q}{2} + \I \R$ or $E_s \in \R$~\cite{Ribault:2015:LiouvilleTheoryCentral}. It was shown numerically in~\cite{Ribault:2015:LiouvilleTheoryCentral} that as long as the internal states belong to $\cS_{\text{int}}$ then the four-point function is well defined and consistent with the full crossing symmetry constraints for external states with any complex momenta.
However, this is not sufficient for our purposes because we would like to define the timelike Liouville theory as an analytic continuation of the theory in~\cite{Ribault:2015:LiouvilleTheoryCentral} in the same way  Lorentzian QFTs are defined by an analytic continuation of Euclidean QFTs. 
That is, the correlation functions with external states in the BRST cohomology must be reached by an appropriate analytic continuation, including the integration contour.

\subsection{Definition of the four-point function}

We restrict ourselves to the task of defining the unintegrated correlation functions of the Liouville sector, because the other (matter and ghosts) sectors are decoupled. There may be additional subtleties in defining the integrated correlation functions but this is a problem for the future. We now give a prescription to define the unintegrated $n$-point functions for $n \ge 4$ for all physical states. To be concrete, let us consider the analytic continuation of the four-point function. Our goal is to define a well-defined crossing-symmetric four-point function for all allowed external states in $\cS_{\text{ext}}$. This can be achieved as follows.
\begin{myenumerate}
	\item
	Start with \eqref{eq:4pt} with all external states with $E_i \in \I \R$, i.e.\  in $\cS_{\text{int}}$. With the choice of the contour along the imaginary $E_{s}$ axis, it is a well-defined crossing-symmetric integral. 
	\item
	As the external energies are analytically continued to $E_i \in \R$, examine if any poles of the integrand cross the contour along the imaginary axis. 
	\item 
	Deform the contour, if necessary, to avoid any of the above poles while holding the ends fixed at $\pm \I \infty$.
\end{myenumerate}

Our prescription is inspired by string field theory (see \S\ref{sec:sft}). 
The main idea is that for an analytic continuation it is not necessary to analytically \textit{rotate} the entire contour as one usually does with the Wick rotation; it is sufficient if one can analytically \textit{deform} the contour while avoiding all poles to reach the physical regime of interest. 

To implement step $(2)$ we need to investigate the poles of the integrand (Figure~\textbf{\ref{fig:poles-4pt}}).
The poles of the conformal block are located at the values of $\alpha_s$ such that the conformal dimension $\Delta_s$ equals the one of a degenerate state, but they are otherwise independent of the external momenta~\cite{Ribault:2014:ConformalFieldTheory}.
Thus these poles do not move as one analytically continues the external energies.
Next, one needs to investigate the poles of the three-point functions. {}From the expression \eqref{eq:3pt-dozz} and the formulas in §\ref{sec:cft:upsilon}, one finds that the poles are located at
\begin{equation}
	\Upsilon_\beta(\beta - 2 \alpha_s) = 0,
\end{equation} 
which are again independent of external energies and hence do not move.

In conclusion, the prescription for timelike gravity is in fact even simpler than the analogous prescription in string field theory. The relevant  poles of the integrand do not move at all as the external energies are analytically continued and it is not even necessary to change the integration contour (which stays the same as in Figure~\textbf{\ref{fig:poles-4pt}}).
The same formula \eqref{eq:4pt} should be used with external states belonging to the $\cS_{\text{ext}}$.
This implies that the code written for~\cite{Ribault:2015:LiouvilleTheoryCentral} is directly usable for the timelike Liouville theory without the need to deform the contour in any way. 
We have explicitly checked numerically that the four-point functions are indeed convergent for all our physical states. 

It was observed already in~\cite{Ribault:2015:LiouvilleTheoryCentral} that the four-point function \eqref{eq:4pt} is analytic in the external momenta and crossing-symmetric for any $\alpha_i \in \C$,\footnote{Using~\cite{Ribault:2015:LiouvilleTheoryCentral}, the correlator appears to be  real when the external states have either all $E\in \R$ or all  $E\in \I\R$.}  and proved that the 1-point function on the torus is modular covariant. 
However, it was assumed in~\cite{Ribault:2015:LiouvilleTheoryCentral} that the spectrum of internal and external states \textit{must be} identical.\footnote{%
	In any case,  the integration over the internal states $p_s \in \R$ has to be shifted by a small real number $p_s \in \R + \I \epsilon$  to avoid poles  located on the imaginary axis.
	Thus, strictly speaking, identifying internal states with  external states is somewhat problematic.
	This point was discussed earlier in §\ref{sec:TLT:spectrum}.
}
From the perspective of quantum gravity, as we have seen, this assumption is unnecessary. The physical spectrum of external states $\cS_{\text{ext}}$ is what it is, and is determined by physical requirements of a ghost-free BRST cohomology. There is no particular physical reason to restrict this spectrum to $\cS_{\text{int}}$ and indeed is not necessary to obtain a meaningful definition of timelike gravity. 

Higher-order correlation functions can be similarly defined by factorization, following for example~\cite{Cho:2017:RecursiveRepresentationsArbitrary}, and analytic continuation from $E \in \I \R$ to $E \in \R$.

\section{Discussion}
\label{sec:discussion}

In this paper, we have provided a definition of the timelike Liouville theory with $c_L \le 1$ by computing its BRST cohomology and proving a no-ghost theorem. We are able to define convergent and crossing-symmetric four-point functions by an appropriate analytic continuation for all external states in the physical spectrum.

Our analytic continuation is motivated by string field theory. 
In that context, there are additional consistency checks on the prescription from the target space. For example, satisfying the Cutkosky rules, proving the unitarity of the theory -- decoupling of unphysical states -- or checking the amplitude crossing symmetry~\cite{Pius:2016:CutkoskyRulesSuperstring, Sen:2016:UnitaritySuperstringField, deLacroix:2018:AnalyticityCrossingSymmetry}. It would be interesting to see if there are similar
consistency checks of our proposal. Regarded as a supercritical string theory, the theory no longer has translation invariance in the time direction because of the cosmological constant. It is thus a time-dependent background with no natural S-matrix interpretation. Therefore, it is not clear what would play the role of S-matrix analyticity~\cite{deLacroix:2018:AnalyticityCrossingSymmetry} in this context.

Much like the critical bosonic string theory, our model regarded as a supercritical bosonic string theory has a tachyon in the target space. This means that at higher genus, integrations over the moduli space will have an infrared divergence from their boundaries. Perhaps this target space infrared divergence should be interpreted as an ultraviolet divergence in the two-dimensional gravity coming from the rapid growth of the density of states at high energies~\cite{Kutasov:1990sv}. Moreover, from the perspective of two-dimensional gravity, higher-genus Riemann surfaces correspond to topology changes.
We do not have much to say about the problem of the tachyons\footnotemark{} or about topology changes.
\footnotetext{%
	Even for $c_L \le 1$, the torus partition function involves only the internal states~\cite{Ribault:2015:LiouvilleTheoryCentral}.
	Since the conformal dimensions of these states are bounded by $\Delta_{\text{min}} = (c_L - 1) / 24$, one finds $c_{\text{eff}} := c_L - 24 \Delta_{\text{min}} = 1$.
	This could help with the tachyons because according to~\cite{Kutasov:1990sv} divergences are expected only if $c_{\text{eff}} > 1$.
}%

Even though our focus was on timelike gravity, our considerations are essentially independent of the matter sector. It can therefore be used also to provide a standalone definition of the timelike Liouville theory as a conformal field theory independent of quantum gravity. Our results suggest that it may be possible to allow the spectrum of external states be larger than the spectrum of internal states even in the CFT context, instead of insisting that the two be the same as was done in~\cite{Ribault:2015:LiouvilleTheoryCentral}. After all, the correlation functions for all these states are well-defined. It would be interesting if there are statistical models which realize this extended CFT. 

Some natural extensions for future work include explorations of the 
supersymmetric versions of this model and models with boundaries
relevant for discussing D-branes and two-dimensional Anti de Sitter space. 
Going back to our original motivation, it would also be interesting to see what general lessons one can draw about the path integral~\cite{Harlow:2011:AnalyticContinuationLiouville} of gravity. 
Finally, an extension of the results for the BRST cohomology for more general matter will be provided in~\cite{Bautista:BRSTCohomologyTimelike}.

\acknowledgments

We thank Dileep Jatkar, Matěj Kudrna, Volker Schomerus, Ashoke Sen, J\"org Teschner  for useful discussions. 
We are particularly grateful to Sylvain Ribault and Raoul Santachiara for very useful discussions, critical comments at various stages of the project and comments on the draft.
The work of H.E.\ is conducted under a Carl Friedrich von Siemens Research Fellowship of the Alexander von Humboldt Foundation.

\appendix

\section{Upsilon function}
\label{sec:cft:upsilon}

\begin{figure}[ht]
	\centering
	\includegraphics[scale=1.5]{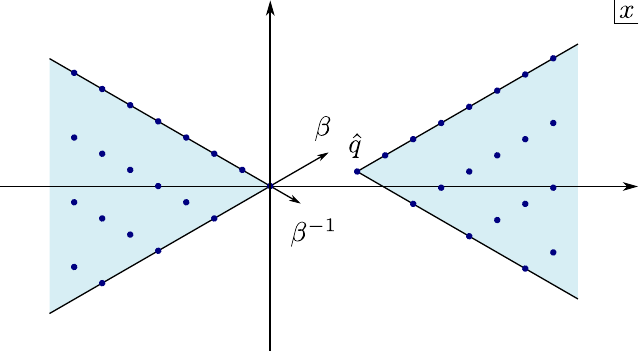}
	\caption{%
		Zeros of the function $\Upsilon_{\b}(x)$ for $\b \in \C$.
		If $\b \in \R$ (resp.\ $\b \in \I \R$), the zeros become all real (resp.\ pure imaginary).
	}%
	\label{fig:zeros-upsilon}
\end{figure}

The $\Upsilon$-function $\Upsilon_\b(x)$ appearing in the three-point correlation functions~\cite{Ribault:2014:ConformalFieldTheory, Dotsenko:2016:AnalyticContinuation3point}
has  a simple integral definition for $\Re( x) \in (0, \Re (\hat q))$:
\begin{equation}
	\ln \Upsilon_\b(x)
		= \int_0^{\infty} \frac{\dd t}{t} \left[
			\left( \frac{\hat q}{2} - x \right)^2 \e^{- 2t}
			- \frac{\sinh^2 \left( \left( \frac{\hat q}{2} - x \right) t\right)}{\sinh (\b t)\, \sinh\left( \frac{t}{\b}\right)}
			\right]
\end{equation} 
where
\begin{equation}
	\hat q = \frac{1}{\b} + \b.
\end{equation} 
This formula admits an analytic continuation to $x \in \C$ and can be represented by an infinite product
\begin{equation}
	\Upsilon_\b(x)
		= \lambda_\b^{\left( \frac{\hat q}{2} - x \right)^2}
			\prod_{m,n \in \N} f\left( \frac{\frac{\hat q}{2} - x}{\frac{\hat q}{2} + m \b + n \b^{-1}} \right),
	\qquad
	f(x) = (1 - x^2) \, \e^{x^2},
\end{equation} 
where $\lambda_\b$ is a constant.
This formula indicates that $\Upsilon_\b(x)$ has no poles and an infinite number of zeros located at (Figure~\ref{fig:zeros-upsilon})
\begin{equation}
	\label{eq:upsilon-zeros}
	(- \b \N - \b^{-1} \N) \cup (\hat q + \b \N + \b^{-1} \N).
\end{equation} 
The function has also a reflection property
\begin{equation}
	\Upsilon_\b(\hat q - x) = \Upsilon_\b(x).
\end{equation} 
A useful limit of this function (to analyse the behaviour of the four-point function integrand) is:
\begin{equation}
	\label{eq:upsilon-large-p}
	\ln \Upsilon_{\b}\left(\frac{\hat q}{2} + \I E \right) \sim_{E \to \infty} - E^2 \ln \abs{E} + \frac{3}{2} \, E^2.
\end{equation}

\section{Comments on the minisuperspace approximation}
\label{sec:minisuperspace}

The minisuperspace approximation has been very useful for studying the spacelike Liouville theory.
However, for the timelike Liouville theory, we were not able  to draw any clear conclusions in this approximation.
We  comment  briefly on a few points encountered in this analysis which will be  discussed further in~\cite{Bautista:BRSTCohomologyTimelike}.

The wave functions for $E \in \R$ cannot be completely fixed by requiring them to decay at infinity, and normalizability does not seem to be sufficient either.
Moreover, contrary to  expectations, the states $E \in \I \R$ are also normalizable and are completely fixed.
Hence, the minisuperspace is not sufficient to completely fix the spectrum as was the case in the spacelike case.
The only possible way to fix the spectrum (up to a parameter taking values in the unit interval) is to perform a self-adjoint extension of the Hamiltonian~\cite{Fulop:1996:ReducedSL2RWZNW, Kobayashi:1996:QuantumMechanicalLiouville, Fredenhagen:2003:MinisuperspaceModelsSbranes, McElgin:2008:NotesLiouvilleTheory}. 
In the latter case, the spectrum consists of  states with $E \in \R$ together with a set of discrete points in $E \in \I\R$, whose interpretation is not clear.
Earlier works~\cite{Strominger:2002:OpenStringCreation, Gutperle:2003:TimelikeBoundaryLiouville, Strominger:2003:CorrelatorsTimelikeBulk, Schomerus:2003:RollingTachyonsLiouville} missed these possibilities by working directly with wave functions analytically continued from the spacelike theory.

Finally, the three-point function computed in the minisuperspace approximation should correspond to the semiclassical limit $\beta \to 0$ of the timelike three-point function \eqref{eq:3pt-dozz} with light states.
However, there is no subset of states in $E \in \R \cup \I\R$ and no constraint on the wave functions such that this is possible. See  appendix \S\ref{sec:history} for more discussion.

\section{A brief history of the timelike structure constant}
\label{sec:history}

Defining the timelike Liouville theory from the action \eqref{sec:TLT} has been the goal of ongoing efforts for 25 years~\cite{Kobayashi:1996:QuantumMechanicalLiouville, Fulop:1996:ReducedSL2RWZNW, Strominger:2002:OpenStringCreation, Strominger:2003:CorrelatorsTimelikeBulk, Gutperle:2003:TimelikeBoundaryLiouville, Schomerus:2003:RollingTachyonsLiouville, Fredenhagen:2003:MinisuperspaceModelsSbranes, Kostov:2004:BoundaryLiouvilleTheory, daCunha:2003:ClosedStringTachyon, Takayanagi:2004:MatrixModelTimelike, McElgin:2008:NotesLiouvilleTheory, Harlow:2011:AnalyticContinuationLiouville, Maltz:2013:GaugeInvariantComputable, Martinec:2014:ModelingQuantumGravity}. Some earlier references to the timelike theory can be found in~\cite{Polchinski:1989:TwodimensionalModelQuantum, Cooper:1991:TwodimensionalQuantumCosmology}).

The identification of the structure constant for the timelike regime has been the subject of several confusions.
In this appendix, we review this history from the modern perspective.

The first difficulty arose from the observation that the structure constant in the minisuperspace approximation takes the form
\begin{equation}
	C_{\text{RW},0}(\alpha_1, \alpha_2, \alpha_3) = H_0(\alpha_1, \alpha_2, \alpha_3)^{-1} \, \hat C_0(\alpha_1, \alpha_2, \alpha_3)
\end{equation} 
where $\hat C_0$ is the expected $\beta \to 0$ limit of the timelike structure constant \eqref{eq:3pt-dozz} for light states
\begin{equation}
	\alpha_i = - \frac{q}{2} + \I \beta E_i,
\end{equation} 
and $H_0$ is non-analytic in the $\alpha_i$. (RW stands for Runkel--Watts-type theories.)
It was found~\cite{Strominger:2003:CorrelatorsTimelikeBulk, Schomerus:2003:RollingTachyonsLiouville, McElgin:2008:NotesLiouvilleTheory} that this structure is reproduced by performing an analytic continuation from $\beta \in \I\R$ to $\beta \in \R$ of the DOZZ formula for $C$, schematically:
\begin{equation}
	C(\I \alpha_1, \I \alpha_2, \I \alpha_3)
		\xrightarrow[\beta \in \R]{}
		C_{\text{RW}}(\alpha_1, \alpha_2, \alpha_3) = H(\alpha_1, \alpha_2, \alpha_3)^{-1} \, \hat C(\alpha_1, \alpha_2, \alpha_3)
\end{equation} 
(see also \S\ref{sec:TLT} for more discussions).
However, this is possible only for the discrete values where $\beta^2 = p / q$ for $p$ and $q$ coprimes (for which $c_L$ takes the same values as for the minimal models).
$C_{\text{RW}}$ looks suspicious to be taken as a definition of the Liouville theory structure constant, since one would have expected it to be defined at all $c \le 1$ and not for discrete values only (which would also restrict severely the matter which can be coupled in the $2d$ gravity context).

This restriction on $\hat C$ comes from insisting on the correctness of the minisuperspace analysis, which for timelike Liouville is not the best guiding principle (as described also in \S\ref{sec:minisuperspace}). 
On the other hand, it has been shown that $\hat C$ as given by \eqref{eq:3pt-dozz} can be reproduced from the path integral by selecting carefully the integration cycle~\cite{Harlow:2011:AnalyticContinuationLiouville} or by Coulomb gas integrals~\cite{Giribet:2012:TimelikeLiouvilleThreepoint}.
Hence, the CFTs defined by the above structure constant $C_{\text{RW}}$ should be interpreted as another family of theories, called the Runkel--Watts models (or non-analytic Liouville in~\cite{Ribault:2015:LiouvilleTheoryCentral}) since the $c = 1$ theory was originally found by Runkel and Watts as a peculiar limit of minimal models~\cite{Runkel:2001:NonrationalCFTC1, Runkel:2002:NonrationalCFTCentral}.
It has been checked in~\cite{Ribault:2015:LiouvilleTheoryCentral} that these models are crossing symmetric.

There was a second motivation for considering the structure constant $C_{\text{RW}}$ instead of $\hat C$.
General expectations from unitary CFTs suggest that the limit $\Delta \to 0$, or $\alpha \to 0$ of a primary operator should yield the identity.
In particular, taking this limit for one external state of the three-point function should give the two-point function.
However, one finds that $\hat C(\alpha_1, 0, \alpha_2)$ does not vanish when $\alpha_1 \neq \alpha_2$, which precludes from identifying it with the two-point function.
This was remarked in~\cite{Strominger:2003:CorrelatorsTimelikeBulk, Schomerus:2003:RollingTachyonsLiouville, McElgin:2008:NotesLiouvilleTheory} and given as a motivation to look for another three-point function which did have the two-point function as the $\alpha \to 0$ limit.
This is indeed the case for $C_{\text{RW}}$, although only for the restricted subset $\beta^2 \in 1 / \N_{\ge 2}$ and after defining the
identity field as a particular limit of the vertex operator~\cite{Runkel:2001:NonrationalCFTC1, Schomerus:2003:RollingTachyonsLiouville, McElgin:2008:NotesLiouvilleTheory}.

The correct interpretation to this issue~\cite{Harlow:2011:AnalyticContinuationLiouville, Ribault:2014:ConformalFieldTheory, Ribault:2015:LiouvilleTheoryCentral, Ikhlef:2016:ThreepointFunctionsc} is that there exists a state with dimension $\Delta = 0$ which is not the identity -- we could call it a “fake identity”.
This can happen in non-unitary CFTs and does not signal any pathology of the theory.
In this case, this state is the highest-weight state of a Verma module instead of the identity degenerate field $1$ (which generates a coset after truncating the null vector and its descendants), and the correlation functions do not satisfy the BPZ equations for this field.
The realization of the $c_L \le 1$ theory in terms of a microscopic loop model~\cite{Ikhlef:2016:ThreepointFunctionsc} gave a concrete geometric identification of this operator.

\section{Analytic continuation in string field theory}
\label{sec:sft}

The divergence of the four-point function integrand \eqref{eq:4-pt-divergence} for $E_s \in \R$ can also be found in loop amplitudes of string field theory (SFT).
Since the detailed construction of SFT is not needed in order to explain how this divergence is cured, we will focus on the relevant information and the reader is referred to the literature (for example the reviews~\cite{Zwiebach:1993:ClosedStringField, deLacroix:2017:ClosedSuperstringField}) for more details.

A string worldsheet amplitude is constructed in the same way as the $2d$ gravity amplitudes, except that the Liouville field is absent in the critical dimension.\footnotemark{}
\footnotetext{%
	Typically, one uses only three vertex with $c$-ghost insertions for tree-level amplitudes: this avoids the introduction of the $b$-ghosts.
	However, this is less natural since the amplitude is not symmetric in the vertex operators and, moreover, the BRST states naturally come with $c$-ghosts~\cite{Polchinski:2005:StringTheory-1}.
}%
The starting point of SFT is to deconstruct the amplitudes in Feynman diagrams by factorization (whose inverse can be understood as a generalization of the OPE)~\cite{Polchinski:2005:StringTheory-1}.
The main interest is to display the infrared divergences of the integrand, for example as two vertex operators come close to each other.

\begin{figure}[ht]
	\centering
	\includegraphics[scale=0.8]{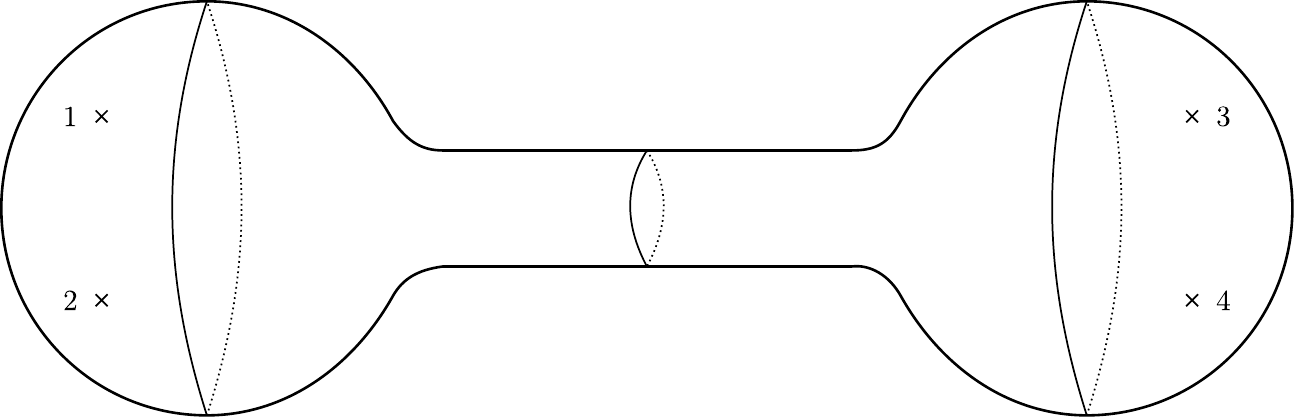}
	\caption{$s$-channel decomposition of the four-point function.}
	\label{fig:4pt-string-ws-s}
\end{figure}

Consider the four-point function as an example: from the spacetime perspective, the limit where the vertex operators $1$ and $2$ approach each other corresponds to an $s$-channel diagram (this corresponds to the boundary of the moduli space).
The associated worldsheet is given in Figure~\ref{fig:4pt-string-ws-s}.
An infrared divergence  would occur when the momenta of $1$ and $2$ are such that the internal particle is on-shell.
To display this divergence explicitly, the standard tree-level four-point function of four string states $A_i$ with momenta $p_\alpha$ (left implicit for simplicity)
\begin{equation}
	G_{0,4} = \int \dd^2 y_4 \, \Mean{\prod_{i=1}^3 \bar c c A_i(z_i) \, A_4(y_4)},
\end{equation} 
where the $z_i$ are fixed, can be rewritten as~\cite{Polchinski:2005:StringTheory-1}
\begin{subequations}
\begin{align}
	G_{0,4}
		&\sim_{z_1 \to z_2} \int \frac{\dd^2 q}{\abs{q}^2} \,
			\Mean{\bar c c V_1(z_1) \bar c c V_2(z_2) \bar b_0 b_0 q^{L_0} \bar{q}^{\bar L_0} \bar c c V_4(z_4) \bar c c V_3(z_3)}
		\\
		\label{eq:string-4pt-factorized}
		&
		\begin{multlined}
		= \sum_{r,s} \int \frac{\dd^D k}{(2\pi)^{D}} \,
			\Mean{\bar c c V_1(z_1) \bar c c V_2(z_2) \phi_r(k)}
			\Mean{\bar c c V_4(z_4) \bar c c V_3(z_3) \phi_s(k)}
			\\
			\times \int \frac{\dd^2 q}{\abs{q}^2} \,
			\Mean{\phi_r^c(k) \bar b_0 b_0 q^{L_0} \bar{q}^{\bar L_0} \phi_s^c(k)}
		\end{multlined}
\end{align}
\end{subequations}
by taking $y_4 = q z_4$ with $z_4$ fixed and performing standard manipulations (due to conformal invariance, the limit $z_1 \to z_2$ is equivalent to $z_4 \to z_3$, moreover, we take $\abs{z_4} < \abs{z_3}$ which explained the chosen radial ordering).
The second formula is obtained by inserting a complete set of states $\{ \phi_r(k) \}$, where $k$ is the $D$-dimensional momentum.
The dual states $\{ \phi_r^c(k) \}$ are defined such that $\bracket{\phi_r^c(k)}{\phi_s(k')} = \delta_{rs} \delta^{(D)}(k - k')$.
From this form, the analogy with the Liouville four-point function \eqref{eq:4pt} should be clear: the first two factors in \eqref{eq:string-4pt-factorized} are the equivalent of the three-point functions, the last factor is equivalent to the conformal block, and the integral over $k$ corresponds to the integration over $\alpha_s$.

Concretely, the single moduli $y_4$ of the $4$-punctured sphere is traded by the variable $q = \e^{- s + \I \theta}$ corresponding to the \emph{plumbing fixture} parameter.
Using this parametrization and performing the integral with $s > 0$ and $\theta \in [0, 2\pi)$ leads to the standard expression
\begin{equation}\label{eqn:propagator-integral}
	\int \frac{\dd^2 q}{\abs{q}^2} \,
			q^{L_0} \bar{q}^{\bar L_0}
		= \int_{0}^{\infty} \dd s \int_0^{2\pi} \dd\theta \,
			\e^{- s (L_0 + \bar L_0)} \e^{\I \theta (L_0 - \bar L_0)}
		= \delta_{L_0, \bar L_0} \, \frac{1}{L_0 + \bar L_0}.
\end{equation} 
Since the basis states $\phi_r^c(k)$ are eigenstates of the $L_0$ and $\bar L_0$ operators, the matrix elements of the propagator match the Feynman propagator
\begin{equation}
	\int \frac{\dd^2 q}{\abs{q}^2} \,
			\Mean{\phi_r^c(k) \bar b_0 b_0 q^{L_0} \bar{q}^{\bar L_0} \phi_s^c(k)}
		= \frac{4 M_{rs}}{k^2 + m_r^2},
\end{equation} 
with some finite-dimensional matrix $M_{rs}$, parametrizing the overlap of states with identical masses.
The parameter $s$ can be interpreted as the length of the intermediate tube in Figure~\ref{fig:4pt-string-ws-s} (Schwinger parameter), while $\theta$ is the twist of the tube.
It should be noted that the parameter $q$ here is not the same as the one in \eqref{eq:4-pt-divergence}: indeed, the plumbing parameter $q$ does not cover the full moduli space, which indicates that one must sum over the $s$-, $t$- and $u$-channels together with a fundamental quartic interaction to reproduce the complete four-point string amplitude.
On the other hand, the elliptic nome maps uniquely the $4$-punctured moduli space.\footnotemark{}
\footnotetext{%
	The reason for separating the moduli space in this way is to make explicit the divergences coming from the moduli space boundaries, which are either spurious or correspond to (target space) IR divergences (which can be treated with renormalization)~\cite{deLacroix:2017:ClosedSuperstringField}.
}%

However, one needs to study loop amplitudes to push the analogy completely: the momentum integral in \eqref{eq:string-4pt-factorized} cannot diverge because the three-point function contains delta functions which remove the integration
\begin{equation}
	\label{eq:expr-propagator}
	\int \dd^D k \, \delta^{(D)}(p_1 + p_2 - k) \delta^{(D)}(p_3 + p_4 - k)
		= \delta^{(D)}(p_1 + p_2 + p_3 + p_4),
\end{equation} 
as it is expected from QFT (see also~\cite[sec.~3.2]{Ribault:2015:LiouvilleTheoryCentral}).
A Feynman diagram for the  general $g$-loop $n$-point Green function $G_{g,n}$ for particles with external momenta $p_\alpha$ is of the form~\cite{Sen:2017:EquivalenceTwoContour, deLacroix:2017:ClosedSuperstringField}
\begin{equation}
	\label{eq:sft-amplitude-gn}
	 \int \dd t \, \int \prod_s \dd^D \ell_s \,
			\e^{- G_{rs}(t) \ell_r \cdot \ell_s - 2 H_{r\alpha}(t) \ell_r \cdot p_\alpha - F_{\alpha\beta}(t) p_\alpha \cdot p_\beta}
			\prod_i \frac{1}{k_i^2 + m_i^2} \,
			\mc P(\ell_r, p_\alpha ; t)\, ,
\end{equation} 
where the $\ell_s$ are the loop momenta, the $k_i = k_i(p_\alpha, \ell_s)$ are the momenta of the internal particles (they are given by a linear combination of the external and loop momenta), $t$ denotes collectively the moduli parameters, and $\mc P$ is a polynomial in the external and loop momenta.
The exponential of the external and loop momenta arises from the OPE between the exponentials in vertex operators.

\begin{figure}[htpb]
	\centering
	\includegraphics[scale=0.8]{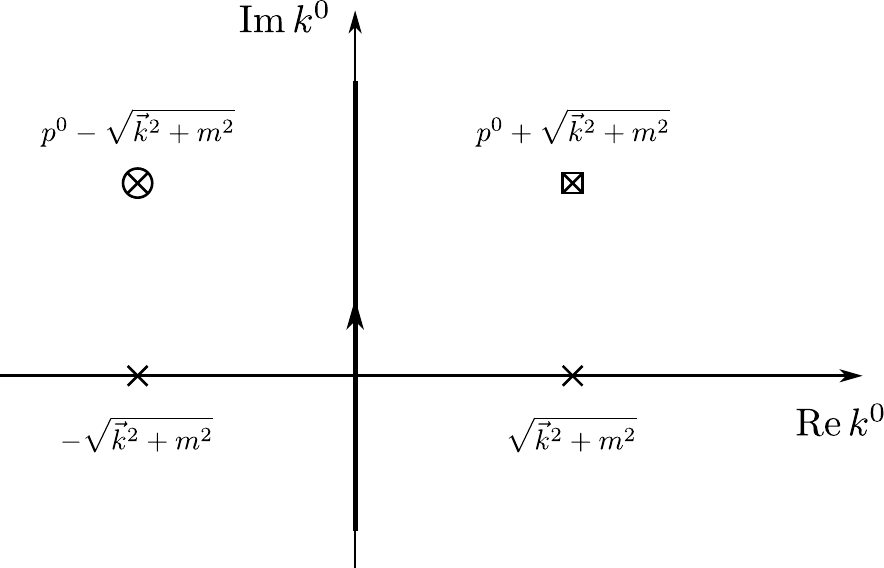}
	\caption{Integration contour for external Euclidean momenta.}
	\label{fig:sft-contour-euclidean}
\end{figure}

\begin{figure}[htpb]
	\centering
	\subcaptionbox{}{\includegraphics[scale=0.8]{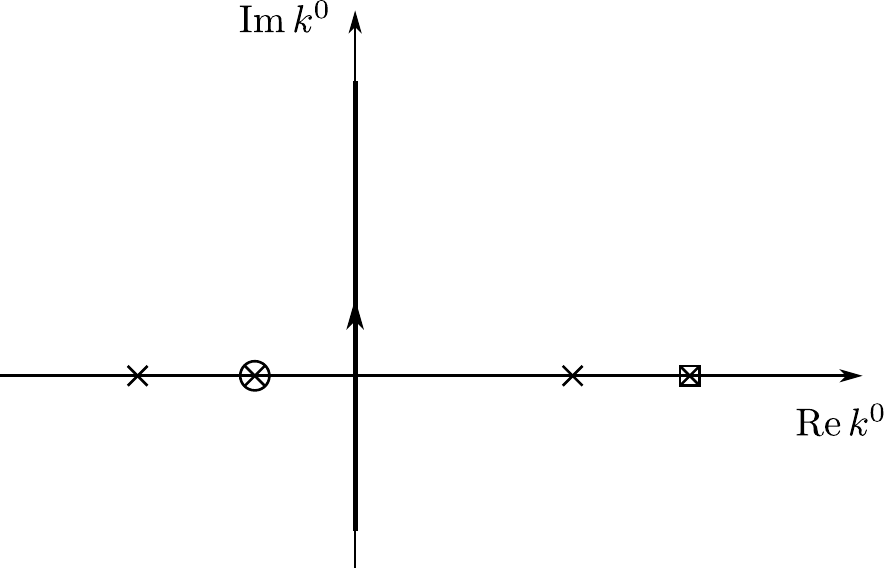}}
	\qquad
	\subcaptionbox{}{\includegraphics[scale=0.8]{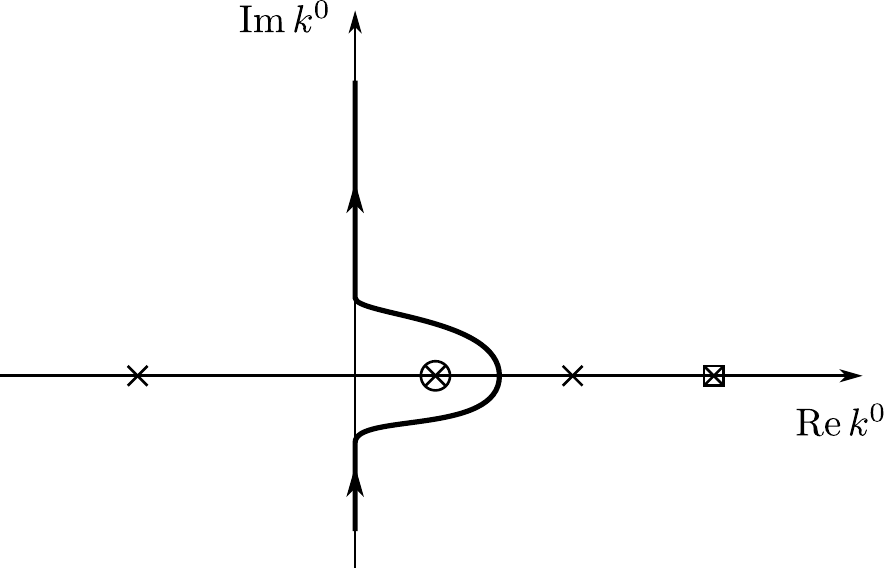}}
	\subcaptionbox{}{\includegraphics[scale=0.8]{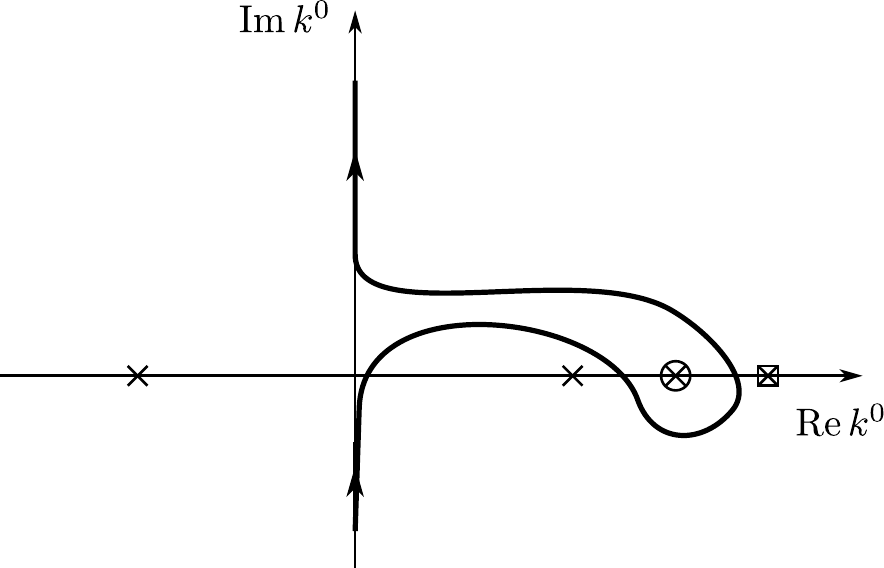}}
	\caption{Integration contour after analytic continuation to external Lorentzian momenta.
		Depending on the values of the external momenta, different cases can happen.
	}
	\label{fig:sft-contour-lorentzian}
\end{figure}

The Feynman diagram \eqref{eq:sft-amplitude-gn}, and as a consequence the amplitude $G_{g,n}$,  has poles when the internal particles become on-shell, i.e.\ for $k_i^2 + m_i^2 = 0$.
Since the matrix $G_{rs}$ is positive-definite (irrespective of $t$), the integrals over the loop momenta converge for the spatial components $\vec \ell_s$, but they diverge for the energy components $\ell_s^0$ (in Minkowski signature)~\cite{Sen:2017:EquivalenceTwoContour, deLacroix:2017:ClosedSuperstringField}.
From the relation \eqref{eqn:propagator-integral} it should now be clear that this divergence is similar to the one encountered in the timelike Liouville theory, equation \eqref{eq:4-pt-divergence}.

In~\cite{Pius:2016:CutkoskyRulesSuperstring}, the SFT amplitudes are defined with the following prescription:
\begin{myenumerate}
	\item Multiply the external energies with a parameter $u \in \C$, such that the external states have momenta $(u\, p_\alpha^0, \vec p_\alpha)$ (Figure~\ref{fig:sft-contour-euclidean}).
	\item Define the amplitudes for Euclidean momenta: $u = \I$ and $\ell_r^0 \in \I \R$.
	The poles of the propagators in \eqref{eq:sft-amplitude-gn} lie at complex values and the integral converges.
	\item Perform an analytic continuation of $u \to 1$ and of the loop momenta $\ell_r^0$, but keeping the end points of the contour at $\pm \I \infty$ and deforming the contour such that each pole remains on the same side 
(Figure~\ref{fig:sft-contour-lorentzian}).
\end{myenumerate}
In this prescription, the internal and external states are different as CFT states (since their momenta lie in different regions).
It has been shown~\cite{Pius:2016:CutkoskyRulesSuperstring, Sen:2016:UnitaritySuperstringField, Sen:2017:EquivalenceTwoContour, deLacroix:2017:ClosedSuperstringField, deLacroix:2018:AnalyticityCrossingSymmetry} that this prescription correctly leads to the Cutkosky rules, unitarity, crossing symmetry of the amplitudes, and general analyticity properties also found in QFT, and that it is equivalent to the moduli space $\I\epsilon$ prescription from~\cite{Witten:2013:FeynmaniEpsilon}.
Since these properties are shared by any good QFT, this provides a strong support for the correctness of the prescription.

\bibliographystyle{JHEP}
\bibliography{timelike.bib}

\end{document}